\documentclass[3p]{elsarticle}
\makeatletter \def\ps@pprintTitle{  \let\@oddhead\@empty  \let\@evenhead\@empty  \def\@oddfoot{\hfill\thepage}  \def\@evenfoot{\thepage\hfill}} \makeatother
\usepackage[T1]{fontenc}
\usepackage{mathtools}
\usepackage{graphicx, color}
\usepackage{hyperref}
\usepackage{natbib}
\usepackage[figuresleft]{rotating}
\usepackage{caption}
\usepackage{subcaption}
\usepackage{layout}
\usepackage{setspace}
\usepackage{lscape}
\usepackage{dcolumn}
\usepackage{booktabs}
\usepackage[misc]{ifsym}
\usepackage{autonum}
\usepackage{multirow}
\usepackage{threeparttablex}
\usepackage{array}
\usepackage{tabularx}
\usepackage{longtable}
\usepackage{commath}
\usepackage{comment}
\usepackage[table]{xcolor}

\biboptions{authoryear}

\journal{}
\begin{document}
\begin{frontmatter}
\title{
On estimating Armington elasticities for Japan's meat imports
}
\author{Satoshi Nakano and Kazuhiko Nishimura }
\cortext[cor1]{Replication data for this study are available at \citet{oniku}. 
}
\begin{abstract}
By fully accounting for the distinct tariff regimes levied on imported meat, we estimate substitution elasticities of Japan's two-stage import aggregation functions for beef, chicken and pork. 
While the regression analysis crucially depends on the price that
consumers face, the post-tariff price of imported meat depends not
only on ad valorem duties but also on tariff rate quotas and gate price system regimes.  
The effective tariff rate is consequently evaluated by utilizing
monthly transaction data. To address potential endogeneity problems, we apply exchange rates that we believe to be independent of the demand shocks for imported meat. 
The panel nature of the data allows us to retrieve the first-stage aggregates via time dummy variables, free of demand shocks, to be used as part of the explanatory variable and as an instrument in the second-stage regression.
\end{abstract}
\begin{keyword}
Two-stage CES aggregation \sep Armington elasticity \sep  Instrumental variables \sep Exchange rates \sep Tariff rate quotas \sep Gate price system
\end{keyword}
\end{frontmatter}

\section{Introduction}
This study focuses on estimating Japan's elasticity of substitution among commodities from different countries (or Armington elasticity) for three different kinds of meat, i.e., beef, chicken and pork.
Our interest is driven by the fact that Japan has been the world's largest meat importer.
The importance of Armington elasticities in the quantitative analysis of various trade policies has been well documented \citep{hh, baj2020}.
We are aware that the welfare implications of any trade policy cannot
be properly evaluated without a finely tuned model with reliable elasticity parameters.
Conversely, reliable estimates of elasticity parameters cannot be obtained unless we are able to properly incorporate the myriad trade regimes that exist. 
In Japan, meat imports are subject to a TRQ (tariff rate quota) regime that allows a lower rate for under-quota imports. 
Pork imports are specifically subject to a GPS (gate price system) regime that discourages imports with prices lower than the gate price.

The main concern of previous related studies has been the difficulty of identifying demand and supply parameters.
The potential simultaneity of demand and supply equations creates
endogeneity problems in any attempt to estimate the elasticity of substitution of imports (i.e., the demand side) by way of a single demand-side equation.
A method established by \citet{feenstra_aer94} and its extensions by \citet{BW_qje06, soderbery_jie15, feenstra_restat18} focus on the combined quadratic equation that delivers consistent estimators under the orthogonality assumption between the demand and supply shocks in a panel data setting.
Feenstra's method is a convenient workaround when one cannot find a
relevant instrument for the endogenous variable (i.e., the post-tariff price).
An orthodox approach to remediate such bias is to find and apply an
instrument that is expected to be independent of the demand shocks.
In this regard \citet{qjz} use tariff rates, while \citet{cje02} use exchange rates.

For the purposes of this study, we seek to evaluate the applied tariff
rates as accurately as possible by fully accounting for the modifications of TRQ regimes and schedule changes of the thresholds governing the GPS. 
The tariff duty under a TRQ depends on in- and out-quota duty rates
and on the cumulative volume of registered imports that determines
which of the two rates applies to each import incident.
To determine the timing of the cumulative volume exceeding the
annually scheduled quotas, we utilize monthly data on import incidents in all cases.
Under a GPS, tariffs are levied to hold the post-tariff price at a  constant level for all import incidents with pretax prices lower than the gate price.
From another perspective, the applied tariff duties under TRQ and GPS
depend on volumes and pretax prices of import incidents, and hence,
they must be correlated with import demand and hence with import demand shocks.
We must therefore rule tariffs out as a potential instrument.

Consequently, we utilize exchange rates to instrument for the endogenous explanatory variable, i.e., the post-tariff price, in the first-stage regression under the assumption that exchange rates and meat import demand shocks are independent.
Our instrument is sufficiently relevant in all cases.
The first-stage regression is based on a multi-input CES function
where the elasticity of substitution (microelasticity) can be
estimated via fixed effects (FE) estimation based on our country-level import observations in time series.  
Additionally, by using time dummy variables, we are able to retrieve the first-stage aggregates from the dummy coefficients and the microelasticity.
Since the estimates of the first-stage aggregates do not contain (the estimates of) the demand shocks, we are able to apply them as an instrument to address the endogeneity in the second-stage regression, where the error term may contain the first-stage demand shocks.
In this way, the second-stage elasticity of substitution (macroelasticity) is estimated.

The remainder of this paper proceeds as follows. 
In the following section, we introduce the two-stage CES aggregation model, deriving two (first- and second-stage) regression equations for estimating the microelasticity and the macroelasticity.
While the microelasticity and first-stage aggregates are estimated by the first-stage regression, the second-stage regression utilizes the first-stage aggregates and estimates the macroelasticity.
In Section 3, we present how we prepare the data for the abovementioned regression analyses.
All applied tariff rates are calculated according to the tariff scheme applied for meat imports to Japan, which we summarize in the Appendix.
Our main results are presented in Section 4, where we show the final estimates of microelasticities and macroelasticities for beef, chicken, and pork.
Section 5 concludes the paper.

\section{Model}
\subsection{Two-stage CES aggregation}
Consider, for some commodity $m$ (index suppressed), a two-stage Armington aggregator as follows:
\begin{align}
{u} = \left( \beta^{\frac{1}{\rho}} {z}^{\frac{\rho-1}{\rho}}+ (1-\beta)^{\frac{1}{\rho}} {y}^{\frac{\rho-1}{\rho}} \right)^{\frac{\rho}{\rho-1}}
&&
{y} = \left( \sum_{i=1}^N \left(\alpha_i\right)^{\frac{1}{\sigma}} \left({x}_i \right)^{\frac{\sigma-1}{\sigma}}\right)^{\frac{\sigma}{\sigma -1}}
\end{align}
where $x_i$ denotes the quantity (of commodity $m$) imported from country $i$, $y$ denotes the utility of aggregated imports, $z$ denotes the quantity produced and consumed in the home country, and $u$ denotes the representative utility in the home country.
Regarding the parameters, $\sigma$ denotes the elasticity of substitution among imports from different countries (or microelasticity), $\rho$ denotes the elasticity of substitution between domestic and aggregate imports (or macroelasticity), and $\alpha_i \geq 0$ and $\beta \geq 0$ are the preference parameters with $\sum_{i=1}^N \alpha_i =1$ and $\beta \leq 1$.
The first function (on the right) is called the first-stage aggregator, and the second (on the left) is called the second-stage aggregator.

The dual function of this two-stage Armington aggregator can be written as follows:
\begin{align}
{v} = \left( \beta r^{1-\rho} + (1- \beta) q^{1-\rho} \right)^{\frac{1}{1-\rho}}
&&
q=\left( \sum_{i=1}^N \alpha_{i}\left(p_{i} \right)^{1-\sigma} \right)^{\frac{1}{1-\sigma}}
\label{2sagg}
\end{align}
where ${p}_i$ denotes the (tariff-inclusive) import commodity price from country $i$ in the home country.
Note that the price of the commodity from the $i$th country $p_i$ (JPY/kg) in terms of Japan's currency unit, domestic price $r$ (JPY/kg), import physical quantity $x_i$ (kg), and domestic physical quantity $z$ (kg) are all observable, but the aggregated values, namely, $y$ (utility), $q$ (JPY/utility), $u$ (utility) and $v$ (JPY/utility), are not.
As per duality, however, we know that the following identities must hold.
\begin{align}
vu = rz + qy
&&
qy = \sum_{i=1}^N p_i x_i
\label{vu}
\end{align}

\subsection{First-stage estimation}
Applying Shephard's lemma for the first-stage aggregator (\ref{2sagg}) yields the following:
\begin{align}
s_i = \frac{p_{it} x_{it} }{\sum_{i=1}^N p_{it} x_{it}}
= \frac{p_i x_i }{qy}
= \frac{\partial q}{\partial p_{i}} \frac{p_{i}}{q} 
= \alpha_{i} \left( \frac{p_{i}}{ q} \right)^{1-\sigma}
\label{snt}
\end{align}
Here, ${s}_i$ denotes the value share of imports of the commodity from the $i$th country.
As we label observations by $t=1,\cdots,J$, we have the following regression equation:
\begin{align}
\ln {s}_{it}
= - (1-\sigma) \ln q_{t} + (1-\sigma) \ln p_{it} +\ln \alpha_{i} +{\epsilon}_{it}
\label{first}
\end{align}
where the error terms ${\epsilon}_{it}$ are assumed to be iid normally distributed with mean zero.
The regression equation (\ref{first}) can be estimated (for $\sigma$ and $q_t$ from $p_{it}$ and $s_{it}$) by FE panel regression.
That is, the first-stage aggregates $q_t$, in terms of indices, are
indirectly measured by the coefficients on the time dummy variables through the FE panel regression.
Let us rewrite regression equation (\ref{first}) using time dummy variables ($D_\ell =1$ if $t=\ell$ and $D_\ell =0$ otherwise), as follows:
\begin{align}
{S}_{it} 
= \sum_{\ell = 1}^J (\mu_\ell - \mu_J) D_{\ell}
+ \mu_{J} + \gamma {P}_{it} + \ln \alpha_i + \epsilon_{it} 
\label{firstreg}
\end{align}
where $S_{it} = \ln s_{it}$, $P_{it} = \ln p_{it}$, and the coefficients therefore denote that
\begin{align}
\mu_t = - \gamma \ln q_t
&&
\gamma = 1 - \sigma
\end{align}
The first-stage aggregates $q_t$ can be resolved, in terms of an index standardized at $t={J}$, as follows:
\begin{align}
q_t = e^{-(\mu_t - \mu_{J})/\gamma} 
&&
t = 1, \cdots, {J}
\label{qhat}
\end{align}

The first-stage regression (\ref{firstreg}) suffers from endogeneity
problems because the demand shock $\epsilon_{it}$ enters the
explanatory variable $P_{it}$ via the potential supply function
connecting $S_{it}$ (the demand for meat from country $i$) to
$\mathcal{F}_{it}$ (the FOB (free on board) price of meat from country $i$), where $\mathcal{F}_{it}$ is a component of $P_{it}$, such that
\begin{align}
{P}_{it} = {C}_{it} + {T}_{it} = (\mathcal{F}_{it} + {E}_{it} + {\Delta}_{it}) + {T}_{it}
\label{decomp}
\end{align}
where $C$ denotes the CIF (cost, insurance and freight) price in JPY
(Japanese yen), $T$ denotes the tariff rate, and $E$ denotes the exchange rate, all in log terms, and ${\Delta}=C -(\mathcal{F} + E)$ denotes the CIF/FOB discrepancy.\footnote{Note that $C$, $T$, and $E$ are observable, while $\mathcal{F}$ and $\Delta$ are not.}
To obtain a consistent estimation for (\ref{firstreg}), we apply exchange rate $E_{it}$ as an instrumental variable for the endogenous explanatory variable $P_{it}$.
Because we believe that demand shocks $\epsilon_{it}$ for meat from country $i$ and the exchange rate $E_{it}$ against the currency of country $i$ are independent, our instrument $E_{it}$ must be exogenous.
Moreover, because $E_{it}$ is a component of $P_{it}$, our instrument
$E_{it}$ must be relevant, i.e., strongly correlated with our
explanatory variable $P_{it}$.\footnote{On the other hand, the
  effective tariff rate $T_{it}$ will not be exogenous because
  $T_{it}$ will depend on quantity demanded $x_{it}$ (which will inevitably be correlated with the demand shock $\epsilon_{it}$) under the tariff regime of import quotas and GPSs.}

\subsection{Second-stage estimation}
Application of Shephard's lemma for the second-stage aggregator of (\ref{2sagg}) yields the following:
\begin{align}
\frac{r}{v}\frac{\partial v}{\partial r} = \frac{rz}{vu} = \beta \left( \frac{r}{v} \right)^{1-\rho} 
&&
\frac{q}{v}\frac{\partial v}{\partial q} = \frac{qy}{vu} = (1-\beta) \left( \frac{q}{v} \right)^{1-\rho} 
\end{align}
By combining the above two identities and labelling observations by $t=1,\cdots, J$, we have the following simple regression equation:
\begin{align}
\ln \left( \frac{r_t z_t}{\sum_{i=1}^N p_{it}x_{it}}\right)
= \ln \left( \frac{\beta}{1-\beta} \right) + (1-\rho) \ln \left( \frac{r_t}{q_t} \right) + \nu_{t}
\label{second}
\end{align}
where $\nu_{t}$ denotes the demand shocks that include the demand shocks for foreign meat ($\epsilon_{it}$) and those for domestic meat ($\delta_{t}$).
The explanatory variable $\ln(r_t/q_t)$, however, must be correlated with $\nu_{t}$ because of the reverse causality of the response variable on the explanatory variable via the possible supply function.

Thus, the second-stage regression (\ref{second}) also suffers from an endogeneity problem.
To obtain consistent estimates, we apply $Q_{t}= \ln \hat{q}_{t}$ to the endogenous explanatory variable $(R_{t} - Q_{t})$ in regression (\ref{second}) which can be rewritten as follows:
\begin{align}
{H}_{t} = \phi + \eta ({R}_{t} - {Q}_{t}) + \nu_{t}
\label{secondreg}
\end{align}
where $\eta = 1 - \rho$, $R_{t} = \ln {r}_{t}$, and so forth.
We suppose that ${Q}_{t}=\ln \hat{q}_{t}$ is exogenous because $\hat{q}_{t}$ does not contain the (estimate of) demand shocks $\hat{\epsilon}_{it}$ and $\delta_{t}$, both of which constitute $\nu_{t}$. 
Moreover, because $Q_{t}$ is a component of the explanatory variable, our instrument $Q_{t}$ must be relevant for our explanatory variable  $(R_{t} - Q_{t})$.

\section{Data compilation}
\subsection{First-stage estimation}
We draw our main data, i.e., monthly import values and quantities from January 1996 to December 2020 for all 78 items whose HS codes are specified in Table \ref{tab_quotas}, from the Commodity by Country link of \cite{boeki}.
Let $v_{it}^g$ and $x_{it}^g$ denote the JPY value and kg quantity of
item $g=1,\cdots, G$ imported from country $i=1,\cdots, N$ at time
$t=1,\cdots, J$, respectively,  where $N=86$, $G=78$, and ${J} = 300$.
Additionally, let $G_m$ denote the set of item IDs of meat type $m$
where $m=1$ denotes beef, $m=2$ denotes pork, and $m=3$ denotes chicken.
Specifically, $G_1 =\left\{1, \cdots,  16 \right\}$, $G_2 =\left\{28, \cdots,  48 \right\}$, and $G_3 =\left\{68, \cdots,  74 \right\}$ as regards to Table \ref{tab_quotas}.
We first prepare the data as follows:
\begin{align}
v_{it} = \sum_{g \in G_m} v_{it}^g
&&
x_{it} = \sum_{g \in G_m} x_{it}^g
&&
c_{it} = \frac{v_{it}}{x_{it}}
\label{cif}
\end{align}
for three kinds of meat ($m=1,2,3$).
Here, $c_{it}$ denotes the CIF price (for one kind of meat) where $\ln
c_{it} = C_{it}$ as mentioned in (\ref{decomp}).
Furthermore, note that $s_{it} = e^{S_{it}}$ is calculated by using $p_{it} = e^{P_{it}}$ from (\ref{decomp}) and $x_{it}$ given above, according to its definition.  

Effective tariff rates $T_{it}$ are evaluated according to Japan's tariff scheme summarized in \ref{appdx}.
We obtain the tariff schedules applied to each item ($g$) with respect to each regime classification from Chapter 2 (Meat and edible meat offal) and Chapter 16 (Preparation of meat etc.) links of \cite{kanzei}.
The above source however only provides tariff schedules from April 2007 onward, and so we refer to the version in Japanese that covers schedules from January 2003 onward.
For earlier schedules prior to 2002, we refer to the hardcopy version
of the Customs Tariff Schedules of Japan (published by the Japan Tariff Association).
We assign the proper tariff schedule to each partner country $i$ in each period $t$ according to the tariff regime classifications available from \cite{wto} (by selecting Japan in the Reporter window and 02 and 16 in the Products window).
The tariff quota schemes (i.e., quota eligible items and annual in-quota quantities) for each bilateral FTA are obtained from the information obtained from \cite{maffquota}.
The monthly in-quota tariff duty for each item ($g$) from each FTA
country ($i$) is evaluated with respect to the monthly cumulative kg quantity of items specified in Table \ref{tab_quotas}. 
Finally, note that historical JPY/LCC exchange rates $e^{E_{it}}$ are drawn from \cite{fx}.

\subsection{Second-stage estimation}
Yearly domestic production data (in tons) from 1996 to 2020 are drawn
from the \cite{estat} Statistical Survey on Livestock (Chikusanbutsu Ryutsu Chosa) for Beef and Pork.
Similar data for chicken are drawn from the \cite{estat} Survey on Broiler Slaughterhouses (Shokucho Shorijo Chosa).
Yearly domestic price data (in JPY/weight) from 1996 to 2020 are drawn
from the \cite{estat} Central Wholesale Meat Market Prices (Shokuniku
Chuo Oroshiuri Shijo Kakaku) for Beef and Pork, and similar data for
chicken are drawn from the \cite{boj} Corporate Goods Price Index (2015 base)/Producer Price Index/Chicken/1995-2020 (item index PR01'PRCG15\_2202050011). 

The abovementioned data sources all provide annual statistics, so our macroelasticity estimation (\ref{secondreg}) must be performed on 25 (1996-2020) observations.
Let us hereafter denote annual periods by $k=1,\cdots,25$.  
The acquired prices and quantities of domestic commodities can then be written as $r_k$ and $z_k$, respectively.
Below, we rewrite the second-stage regression (\ref{secondreg}) for annual observations:
\begin{align}
{H}_{k} = \phi + \eta ({R}_{k} - {Q}_{k}) + \nu_{k}
\label{secondreg2}
\end{align}
where ${R}_{k} = \ln r_k$. 
Let $J(k)$ denote a set of monthly periods ($t$) within year $k$.
We can then aggregate our monthly variables into annual variables in the following manner:
\begin{align}
\hat{q}_{k} = \frac{\sum_{t \in J(k)} w_t}{\sum_{t \in J(k)} w_t/\hat{q}_t}
&&
h_k = \frac{r_k z_k}{\sum_{i=1}^N w_{ik}}
\end{align}
where we define $w_t = \sum_{i=1}^N p_{it} x_{it}$, and $w_{ik} =\sum_{t \in J(k)} p_{it} x_{it}$. 
In this way, our variables $H_k = \ln h_k$ and $Q_k = \ln \hat{q}_k$ for the second-stage regression (\ref{secondreg2}) are prepared. 

\section{Results}
\subsection{First-stage estimation for beef and chicken}
After the data preparation process described above, we perform panel regression analysis based on (\ref{firstreg}) to estimate microelasticities and obtain first-stage aggregates via (\ref{qhat}).
The results for beef are presented in Table \ref{tab_1} (for microelasticity) and Figure \ref{fig_BandC} left (for first-stage aggregates), and those for chicken are presented in Table \ref{tab_2} (for microelasticity) and Figure \ref{fig_BandC} right (for first-stage aggregates).
We apply a heteroskedasticity- and autocorrelation-consistent estimator based on Bartlett's kernel with bandwidth set equal to $5 \approx T^(1/4) +1$, in all first-stage estimations.  
Moreover, for the sake of credibility, we drop panels that have fewer than or equal to 9 observations of the 300 possible time periods spanning 25 years.
\newcolumntype{.}{D{.}{.}{3}}
\begin{table}[t!]
\caption{First-stage estimation for beef} \label{tab_1}\vspace{-10pt}
\begin{center}
\begin{threeparttable}
\begin{tabular*}{0.95\textwidth}{@{\extracolsep{\fill}}lccccrrr}																									
\hline\noalign{\smallskip}																									
\multicolumn{1}{l}{}&\multicolumn{2}{c}{FE (LS)}&\multicolumn{2}{c}{FE (IV)}&\multicolumn{3}{c}{Delta Method}\\\cmidrule(r){2-3}\cmidrule(r){4-5}\cmidrule(r){6-8}																									
&\multicolumn{1}{c}{coef.}&\multicolumn{1}{c}{s.e.}&\multicolumn{1}{c}{coef.}&\multicolumn{1}{c}{s.e.}&&\multicolumn{1}{c}{estim.}&\multicolumn{1}{c}{s.e.}\\\cmidrule(r){1-8}																									
		\multicolumn{1}{l}{	\texttt{lnp}	}&	\multicolumn{1}{.}{	-1.245	}&	\multicolumn{1}{.}{	0.240	}&	\multicolumn{1}{.}{	-3.354	}&	\multicolumn{1}{.}{	0.831	}&	\multicolumn{1}{c}{	$\sigma$	}&	\multicolumn{1}{.}{	4.354	}&	\multicolumn{1}{.}{	0.831	}\\ \cmidrule(r){1-8}
		\multicolumn{1}{l}{	obs.	}&	\multicolumn{2}{c}{	2,501	}&	\multicolumn{2}{c}{	2,406	}&	\multicolumn{3}{l}{		}\\												
\end{tabular*}																									
\begin{tabular*}{0.95\textwidth}{lrrrrrr..}																									
\multicolumn{9}{l}{--- Tests for 2SLS FE (IV) estimation ---\texttt{ lnjpylcc lnyjpylcc}}\\																									
	&	\multicolumn{5}{l}{	Underidentification}{Kleibergen-Paap rk LM statistic							}&	{		}&	\multicolumn{1}{.}{	26.167	}&	\multicolumn{1}{.}{	(0.000)	}\\						
	&	\multicolumn{5}{l}{	Weak identification}{Kleibergen-Paap rk Wald F statistic							}&	{		}&	\multicolumn{1}{.}{	15.341	}&	\multicolumn{1}{.}{		}\\						
	&	\multicolumn{5}{l}{	Overidentifying restriction}{Hansen J statistic							}&	{		}&	\multicolumn{1}{.}{	1.19	}&	\multicolumn{1}{.}{	(0.275)	}\\						
	&	\multicolumn{5}{l}{	Endogeneity}{statistic							}&	{		}&	\multicolumn{1}{.}{	7.885	}&	\multicolumn{1}{.}{	(0.005)	}\\						
\end{tabular*}																									
\begin{tabular*}{0.95\textwidth}{@{\extracolsep{\fill}}l}																									
\hline																									
\end{tabular*}																									
\begin{tablenotes}
\small
\item[Note 1] The numbers in parentheses for all tests are the p-values for rejecting the null hypotheses. 
\item[Note 2] Delta method estimates are based on FE (IV) according to the endogeneity test result.
\item[Note 3] All standard errors are robust to heteroskedasticity and autocorrelation.
\item[Note 4] Applied instruments: 1) log of exchange rates and 2) log of annual cumulated exchange rates.
\end{tablenotes}
\end{threeparttable}
\end{center}
\end{table}
\begin{table}[t!]
\caption{First-stage estimation for chicken} \label{tab_2}\vspace{-10pt}
\begin{center}
\begin{threeparttable}
\begin{tabular*}{0.95\textwidth}{@{\extracolsep{\fill}}lccccrrr}																									
\hline\noalign{\smallskip}																									
\multicolumn{1}{l}{}&\multicolumn{2}{c}{FE (LS)}&\multicolumn{2}{c}{FE (IV)}&\multicolumn{3}{c}{Delta Method}\\\cmidrule(r){2-3}\cmidrule(r){4-5}\cmidrule(r){6-8}																									
&\multicolumn{1}{c}{coef.}&\multicolumn{1}{c}{s.e.}&\multicolumn{1}{c}{coef.}&\multicolumn{1}{c}{s.e.}&&\multicolumn{1}{c}{estim.}&\multicolumn{1}{c}{s.e.}\\\cmidrule(r){1-8}																									
		\multicolumn{1}{l}{	\texttt{lnp}	}&	\multicolumn{1}{.}{	-0.589	}&	\multicolumn{1}{.}{	0.490	}&	\multicolumn{1}{.}{	-3.011	}&	\multicolumn{1}{.}{	0.749	}&	\multicolumn{1}{c}{	$\sigma$	}&	\multicolumn{1}{.}{	4.011	}&	\multicolumn{1}{.}{	0.749	}\\ \cmidrule(r){1-8}
		\multicolumn{1}{l}{	obs.	}&	\multicolumn{2}{c}{	2,834	}&	\multicolumn{2}{c}{	2,834	}&	\multicolumn{3}{l}{		}\\												
\end{tabular*}																									
\begin{tabular*}{0.95\textwidth}{lrrrrrr..}																									
\multicolumn{9}{l}{--- Tests for 2SLS FE (IV) estimation ---\texttt{ jpylcc yjpylcc}}\\																									
	&	\multicolumn{5}{l}{	Underidentification}{Kleibergen-Paap rk LM statistic							}&	{		}&	\multicolumn{1}{.}{	25.294	}&	\multicolumn{1}{.}{	(0.000)	}\\						
	&	\multicolumn{5}{l}{	Weak identification}{Kleibergen-Paap rk Wald F statistic							}&	{		}&	\multicolumn{1}{.}{	19.34	}&	\multicolumn{1}{.}{		}\\						
	&	\multicolumn{5}{l}{	Overidentifying restriction}{Hansen J statistic							}&	{		}&	\multicolumn{1}{.}{	0.004	}&	\multicolumn{1}{.}{	(0.951)	}\\						
	&	\multicolumn{5}{l}{	Endogeneity}{statistic							}&	{		}&	\multicolumn{1}{.}{	8.498	}&	\multicolumn{1}{.}{	(0.004)	}\\						
\end{tabular*}																									
\begin{tabular*}{0.95\textwidth}{@{\extracolsep{\fill}}l}																									
\hline																									
\end{tabular*}																									
\begin{tablenotes}
\small
\item[Note 1] The numbers in parentheses for all tests are the p-values for rejecting the null hypotheses. 
\item[Note 2] Delta method estimates are based on FE (IV) according to the endogeneity test result.
\item[Note 3] All standard errors are robust to heteroskedasticity and autocorrelation.
\item[Note 4] Applied instruments: 1) exchange rates and 2) annual cumulated exchange rates.
\end{tablenotes}
\end{threeparttable}
\end{center}
\end{table}

Let us briefly review the diagnostics regarding the two-stage least squares (2SLS) fixed effects instrumental variables (FE (IV)) estimation.
The first two tests concern whether the instruments (say, $E$) are relevant predictors of the endogenous explanatory variable (say, $P$).
The corresponding statistic for the underidentification test is used to assess the null hypothesis that the minimal canonical correlations between $P$ and $E$ are zero.
The relevance of instruments is further examined by the weak identification test.
The rule of thumb for rejection of the null hypothesis that $E$ is only weakly correlated with $P$ is for the corresponding statistic to exceed 10.
The third test (overidentification restriction) concerns the exogeneity of the (multiple) instruments.
The corresponding statistic examines the null hypothesis that the instruments are \textit{uncorrelated} with the residuals given that at least one of the instruments is exogenous.

The fourth test (endogeneity) is concerned with the endogeneity of the regressor in the FE setting.
A rejection of the null hypothesis indicates that the instrumental variables estimator should be employed over the least squares estimator, in terms of the efficiency of inference.
For both (beef and chicken) cases, based on the endogeneity test
results, FE (IV) is chosen for further inference on the estimates of
microelasticity and first-stage aggregates, via the delta method. 
Final microelasticities estimation results are presented in the Delta Method column of Tables \ref{tab_1} and \ref{tab_2}.
Figure \ref{fig_BandC} depicts the levels of the first-stage aggregates as indices with corresponding standard errors.

\subsection{Second-stage estimation for beef and chicken}
Given the time-series nature of the regression equation (\ref{secondreg2}), we first examine the stationarity of all variables.
The ADF unit root test results on all variables ($H_k$ and $(R_k - Q_k)$) in levels and differences (not reported for brevity) suggest that the variables are stationary in first differences for both cases (beef and chicken).
We also conduct the Engle-Granger cointegration test; the results (not
reported for brevity) suggest that the null hypothesis of no
cointegration would not be rejected for the case of beef, while the
converse would be true for the case of chicken.
We therefore estimate macroelasticities by the first differences of
variables in the case of beef, while we do so by the levels of variables in the case of chicken.
The results are presented in Tables \ref{tab_5} and \ref{tab_6} for beef and chicken, respectively.
For both cases, the first-stage aggregator and its log are applied as instrumental variables for the corresponding explanatory variable.
Note further that we were able to make a proper assessment of the macro-share parameter $\beta$ in the case of chicken in Table \ref{tab_6}.
\begin{figure}[t!]
\centering
\includegraphics[width=0.48\textwidth]{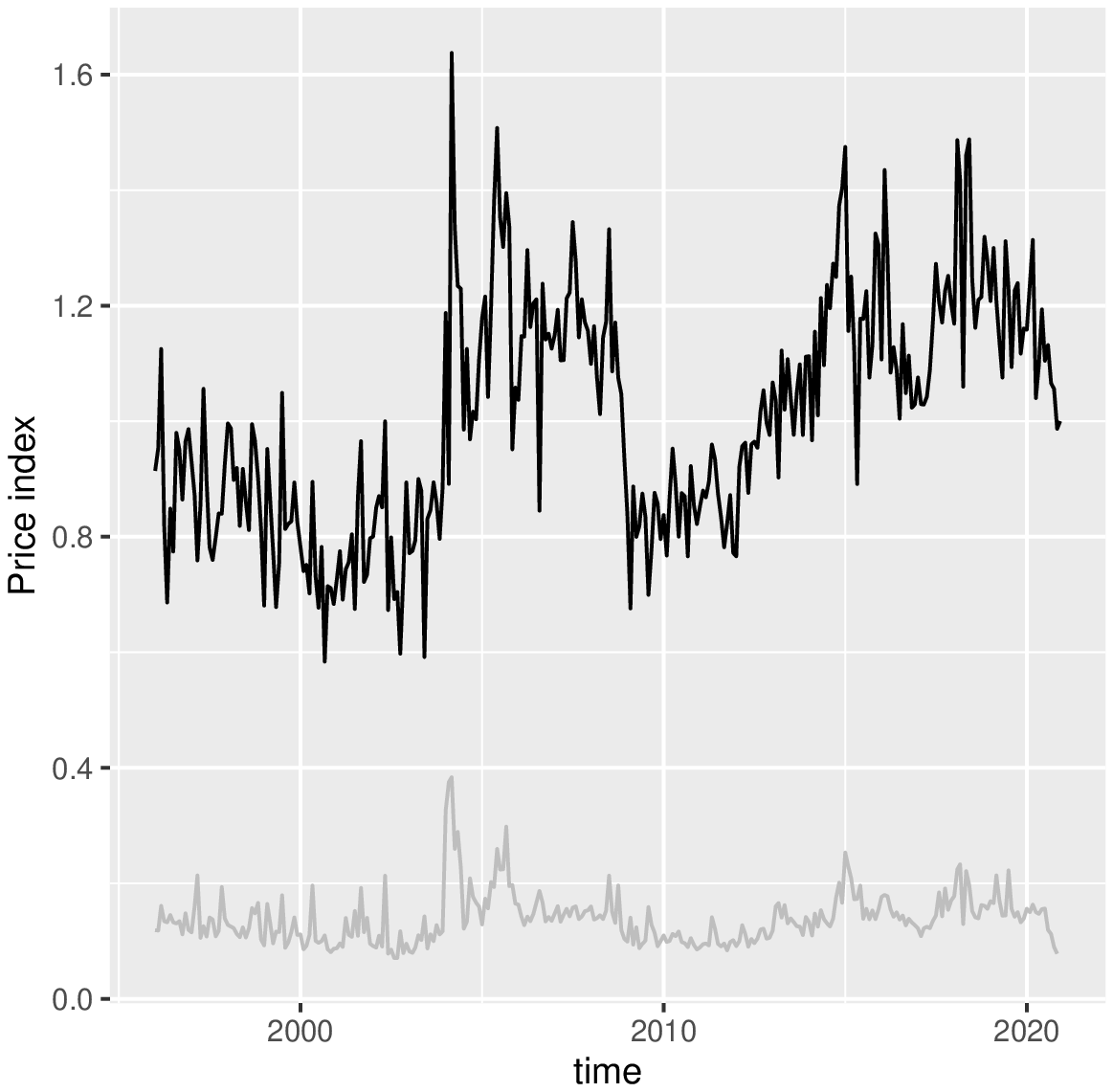}
\includegraphics[width=0.48\textwidth]{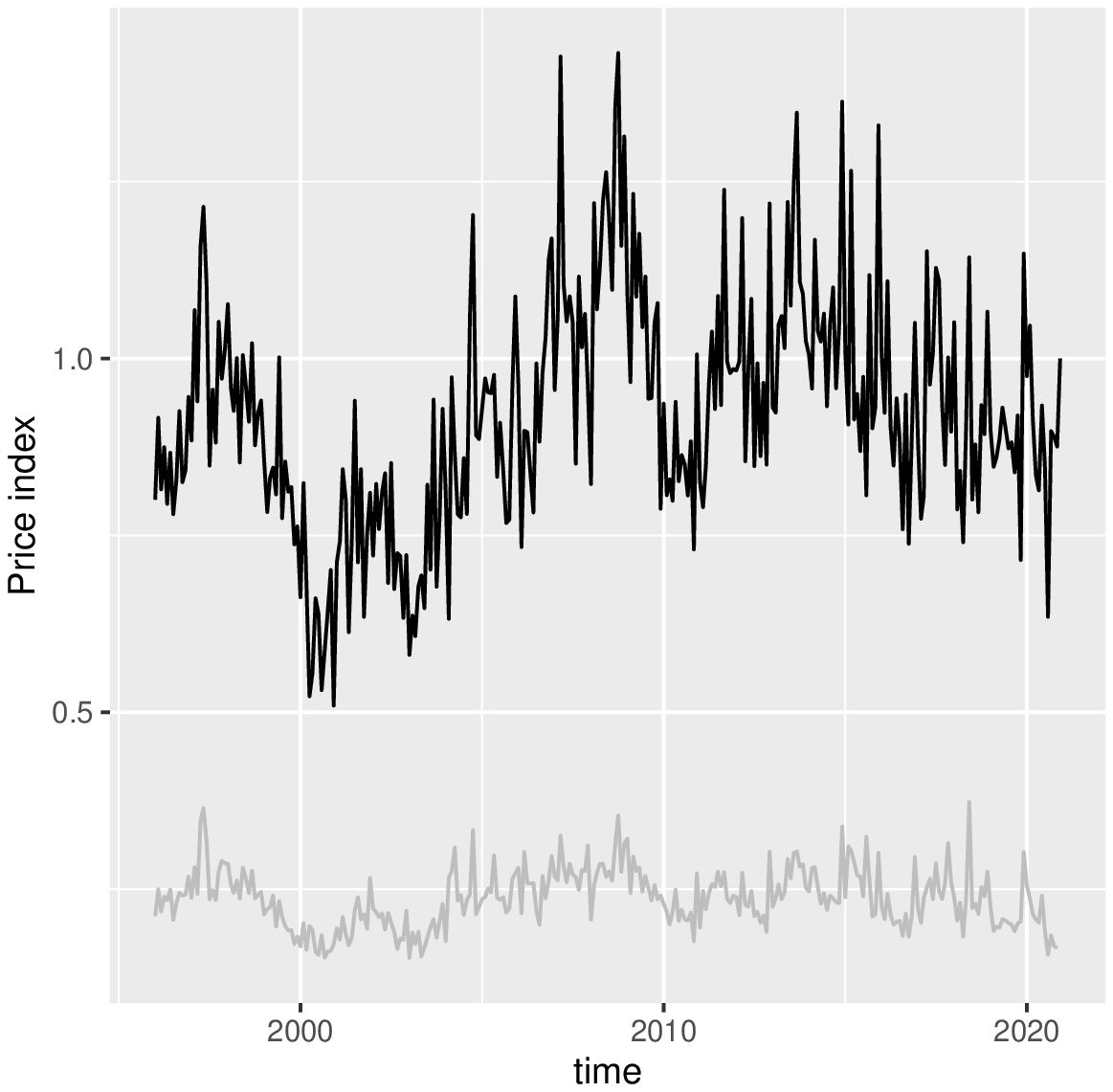}
\caption{
Point estimates of the first-stage aggregates $\hat{q}_t$ (solid line) and their corresponding standard errors (gray line) for beef (left) and chicken (right).
According to (\ref{qhat}), the levels are standardized at the last period $t=J=300=\text{Dec 2020}$.
} \label{fig_BandC}
\end{figure}
\newcolumntype{.}{D{.}{.}{3}}
\begin{table}[t!]
\caption{Second-stage estimation for beef} \label{tab_5}\vspace{-10pt}
\begin{center}
\begin{threeparttable}
\begin{tabular*}{0.95\textwidth}{@{\extracolsep{\fill}}lccccrrr}																									
\hline\noalign{\smallskip}																									
\multicolumn{1}{l}{}&\multicolumn{2}{c}{FE (LS)}&\multicolumn{2}{c}{FE (IV)}&\multicolumn{3}{c}{Delta Method}\\\cmidrule(r){2-3}\cmidrule(r){4-5}\cmidrule(r){6-8}																									
&\multicolumn{1}{c}{coef.}&\multicolumn{1}{c}{s.e.}&\multicolumn{1}{c}{coef.}&\multicolumn{1}{c}{s.e.}&&\multicolumn{1}{c}{estim.}&\multicolumn{1}{c}{s.e.}\\\cmidrule(r){1-8}																									
		\multicolumn{1}{l}{	\texttt{d.lnrq}	}&	\multicolumn{1}{.}{	-0.141	}&	\multicolumn{1}{.}{	0.378	}&	\multicolumn{1}{.}{	-0.001	}&	\multicolumn{1}{.}{	0.527	}&	\multicolumn{1}{c}{	$\rho$	}&	\multicolumn{1}{.}{	1.141	}&	\multicolumn{1}{.}{	0.378	}\\ \cmidrule(r){1-8}
		\multicolumn{1}{l}{	obs.	}&	\multicolumn{2}{c}{	24	}&	\multicolumn{2}{c}{	24	}&	\multicolumn{3}{l}{		}\\												
\end{tabular*}																									
\begin{tabular*}{0.95\textwidth}{lrrrrrr..}																									
\multicolumn{9}{l}{--- Tests for 2SLS FE (IV) estimation ---\texttt{ lnjpylcc lnyjpylcc}}\\																									
	&	\multicolumn{5}{l}{	Underidentification}{Kleibergen-Paap rk LM statistic							}&	{		}&	\multicolumn{1}{.}{	5.391	}&	\multicolumn{1}{.}{	(0.068)	}\\						
	&	\multicolumn{5}{l}{	Weak identification}{Kleibergen-Paap rk Wald F statistic							}&	{		}&	\multicolumn{1}{.}{	55.205	}&	\multicolumn{1}{.}{		}\\						
	&	\multicolumn{5}{l}{	Overidentifying restriction}{Hansen J statistic							}&	{		}&	\multicolumn{1}{.}{	1.272	}&	\multicolumn{1}{.}{	(0.260)	}\\						
	&	\multicolumn{5}{l}{	Endogeneity}{statistic							}&	{		}&	\multicolumn{1}{.}{	0.225	}&	\multicolumn{1}{.}{	(0.635)	}\\						
\end{tabular*}																									
\begin{tabular*}{0.95\textwidth}{@{\extracolsep{\fill}}l}																									
\hline																									
\end{tabular*}																									
\begin{tablenotes}
\small
\item[Note 1] The numbers in parentheses for all tests are the p-values for rejecting the null hypotheses. 
\item[Note 2] Delta method estimates are based on LS according to the endogeneity test result.
\item[Note 3] All standard errors are robust to heteroskedasticity.
\item[Note 4] Applied instruments: first differences of 1) log of first-stage aggregates and 2) first-stage aggregates.
\end{tablenotes}
\end{threeparttable}
\end{center}
\end{table}

\begin{figure}[t!]
\centering
\includegraphics[width=0.48\textwidth]{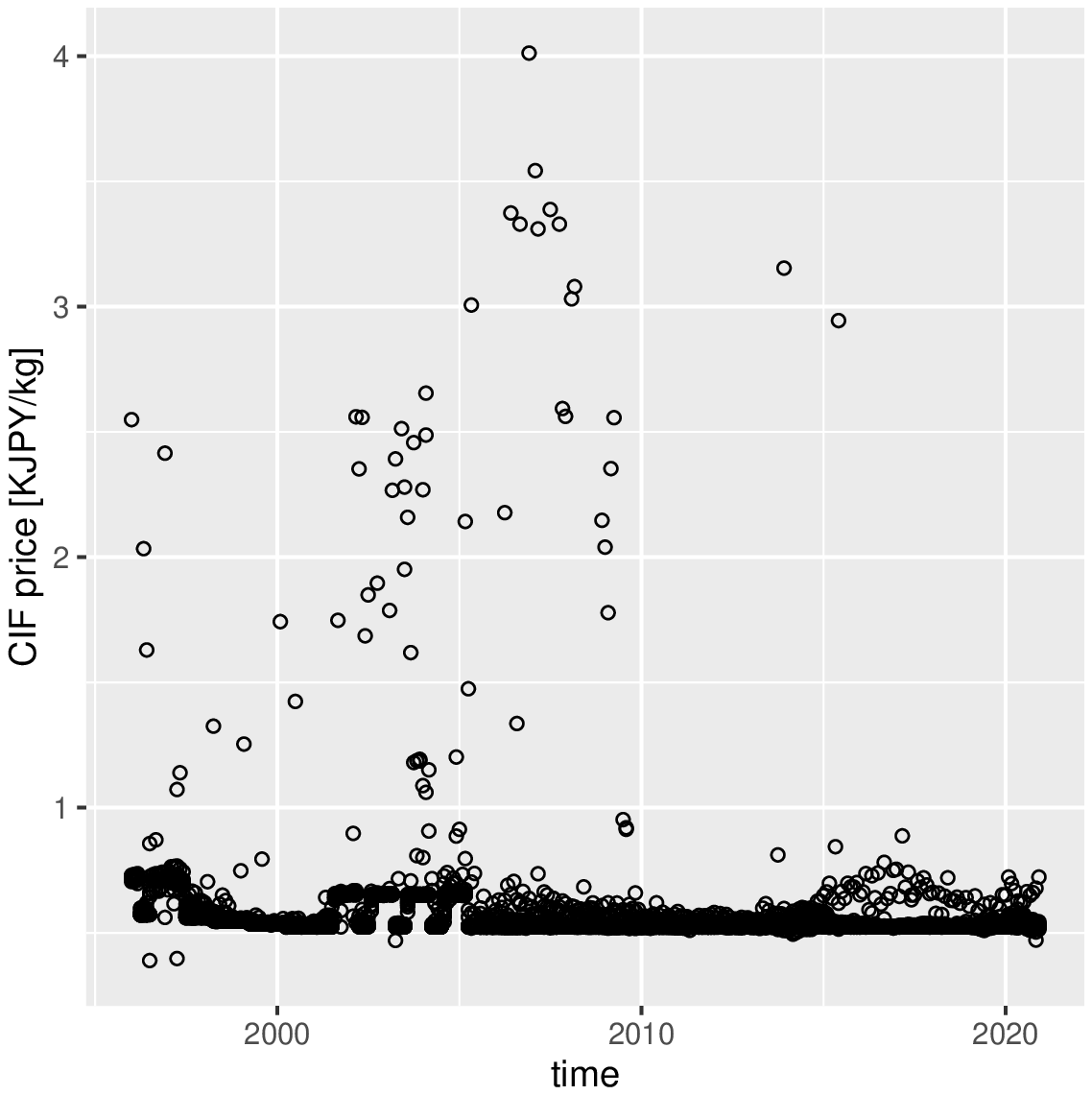}
\includegraphics[width=0.48\textwidth]{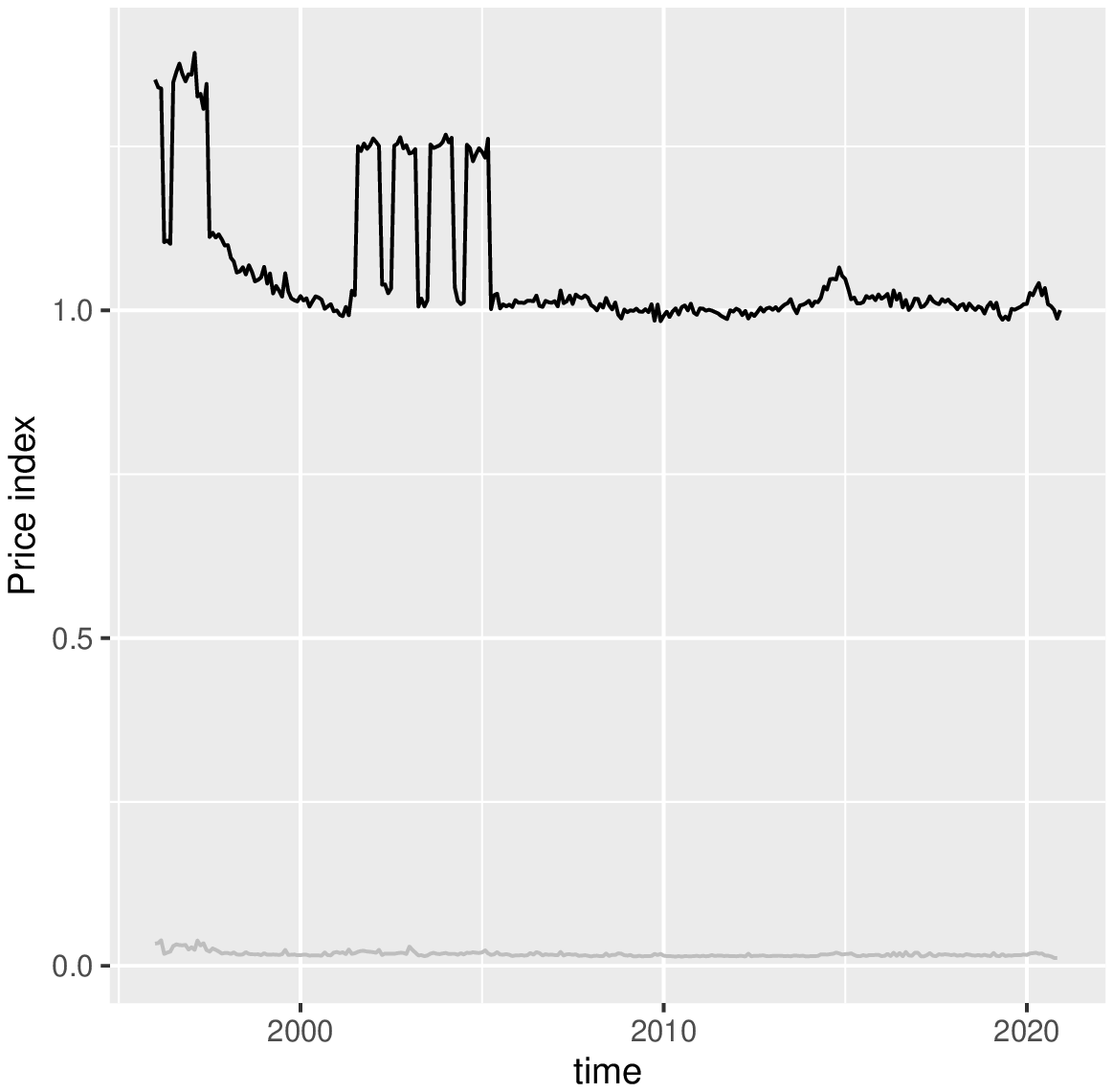}
\caption{
Left: Scatterplot of the CIF price ($c_{it}$) compiled by way of (\ref{cif}) for pork.
Right: Point estimates of the first-stage aggregates $\hat{q}_t$
(solid line) and their corresponding standard errors (gray line) for
pork (regular).
} \label{fig_fig_p}
\end{figure}

\begin{table}[t!]
\caption{Second-stage estimation for chicken} \label{tab_6}\vspace{-10pt}
\begin{center}
\begin{threeparttable}
\begin{tabular*}{0.95\textwidth}{@{\extracolsep{\fill}}lccccrrr}																									
\hline\noalign{\smallskip}																									
\multicolumn{1}{l}{}&\multicolumn{2}{c}{FE (LS)}&\multicolumn{2}{c}{FE (IV)}&\multicolumn{3}{c}{Delta Method}\\\cmidrule(r){2-3}\cmidrule(r){4-5}\cmidrule(r){6-8}																									
&\multicolumn{1}{c}{coef.}&\multicolumn{1}{c}{s.e.}&\multicolumn{1}{c}{coef.}&\multicolumn{1}{c}{s.e.}&&\multicolumn{1}{c}{estim.}&\multicolumn{1}{c}{s.e.}\\\cmidrule(r){1-8}																									
		\multicolumn{1}{l}{	\texttt{lnrq}	}&	\multicolumn{1}{.}{	-0.105	}&	\multicolumn{1}{.}{	0.238	}&	\multicolumn{1}{.}{	-0.088	}&	\multicolumn{1}{.}{	0.200	}&	\multicolumn{1}{c}{	$\rho$	}&	\multicolumn{1}{.}{	1.105	}&	\multicolumn{1}{.}{	0.238	}\\
		\multicolumn{1}{l}{	\texttt{const.}	}&	\multicolumn{1}{.}{	0.367	}&	\multicolumn{1}{.}{	0.034	}&	\multicolumn{1}{.}{	0.367	}&	\multicolumn{1}{.}{	0.034	}&	\multicolumn{1}{c}{	$\beta$	}&	\multicolumn{1}{.}{	0.591	}&	\multicolumn{1}{.}{	0.008	}\\ \cmidrule(r){1-8}
		\multicolumn{1}{l}{	obs.	}&	\multicolumn{2}{c}{	25	}&	\multicolumn{2}{c}{	25	}&	\multicolumn{3}{l}{		}\\												
\end{tabular*}																									
\begin{tabular*}{0.95\textwidth}{lrrrrrr..}																									
\multicolumn{9}{l}{--- Tests for 2SLS FE (IV) estimation ---\texttt{lnpfi pfi}}\\																									
	&	\multicolumn{5}{l}{	Underidentification}{Kleibergen-Paap rk LM statistic							}&	{		}&	\multicolumn{1}{.}{	9.554	}&	\multicolumn{1}{.}{	(0.008)	}\\						
	&	\multicolumn{5}{l}{	Weak identification}{Kleibergen-Paap rk Wald F statistic							}&	{		}&	\multicolumn{1}{.}{	67.765	}&	\multicolumn{1}{.}{		}\\						
	&	\multicolumn{5}{l}{	Overidentifying restriction}{Hansen J statistic							}&	{		}&	\multicolumn{1}{.}{	0.003	}&	\multicolumn{1}{.}{	(0.954)	}\\						
	&	\multicolumn{5}{l}{	Endogeneity}{statistic							}&	{		}&	\multicolumn{1}{.}{	0.017	}&	\multicolumn{1}{.}{	(0.895)	}\\						
\end{tabular*}																									
\begin{tabular*}{0.95\textwidth}{@{\extracolsep{\fill}}l}																									
\hline																									
\end{tabular*}																									
\begin{tablenotes}
\small
\item[Note 1] The numbers in parentheses for all tests are the p-values for rejecting the null hypotheses. 
\item[Note 2] Delta method estimates are based on LS according to the endogeneity test result.
\item[Note 3] All standard errors are robust to heteroskedasticity.
\item[Note 4] Applied instruments: 1) log of first-stage aggregates and 2) first-stage aggregates.
\end{tablenotes}
\end{threeparttable}
\end{center}
\end{table}

\subsection{First- and Second-stage estimations for pork}
Let us first examine the CIF price (${c_{it}}$) we compiled by way of (\ref{cif}) for pork, which is depicted in Figure \ref{fig_fig_p}, left.
This figure clearly demonstrates that the CIF price of pork is affected by the gate price in the timeline, as depicted in Figure \ref{fig_gps}.
Note that most pork imports are priced very near to the gate price where tax payment is minimized.
We thus infer that importers combine expensive meat with inexpensive meat within a container, so that the meat price (of the container) nears the gate price. 
We then closely examine the data composing $c_{it}$ and find that 4730
(98.5\%) observations are under 0.8 KJPY/kg; among the remaining 74
(1.5\%) observations that were over 0.8 KJPY/kg, 72\% came from Italy
and Spain. 

We therefore opt to split the observations into two classes and
conduct the first-stage estimation separately.
For convenience, let us hereafter denote the set of observations whose
CIF prices are under 0.8 KJPY/kg by pork (regular) and its complementary set
(that are more expensive than 0.8 KJPY/kg) by pork (prime).
The results of estimating the microelasticity for pork (regular) based on (\ref{firstreg}) are presented in Table \ref{tab_3}. 
The corresponding point estimates of the first-stage aggregates are depicted in Figure \ref{fig_fig_p}, right, with standard errors.
Note that a large elasticity is consistent with the small variation in
the CIF prices of pork (regular).

In the case of pork (prime), however, our instruments are not relevant.
We then focus on the feedback channel via the potential supply function from $S$ of (\ref{firstreg}) into $\mathcal{F}$ of (\ref{decomp}), which causes the endogeneity problem by letting the error term $\epsilon$ enter the explanatory variable $P$.
If this channel is to be effective, $E$ must also enter $\mathcal{F}$, in which case $E$ and $\mathcal{F} + \Delta = C - E$ become correlated.
Therefore, we regress $E$ on $C-E$ by FE and find that there are no valid interactions between these two variables and, therefore, no channel for endogeneity.
We hence applied FE (LS) to obtain the estimates of the
microelasticity for the case of pork (prime) as shown in Table \ref{tab_4}.
Note however that we could not retrieve the estimates of the first-stage aggregates due to lack of observations within the range of the timeline.

\begin{table}[t!]
\caption{First-stage estimation for pork (regular)} \label{tab_3}\vspace{-10pt}
\begin{center}
\begin{threeparttable}
\begin{tabular*}{0.95\textwidth}{@{\extracolsep{\fill}}lccccrrr}																									
\hline\noalign{\smallskip}																									
\multicolumn{1}{l}{}&\multicolumn{2}{c}{FE (LS)}&\multicolumn{2}{c}{FE (IV)}&\multicolumn{3}{c}{Delta Method}\\\cmidrule(r){2-3}\cmidrule(r){4-5}\cmidrule(r){6-8}																									
&\multicolumn{1}{c}{coef.}&\multicolumn{1}{c}{s.e.}&\multicolumn{1}{c}{coef.}&\multicolumn{1}{c}{s.e.}&&\multicolumn{1}{c}{estim.}&\multicolumn{1}{c}{s.e.}\\\cmidrule(r){1-8}																									
		\multicolumn{1}{l}{	\texttt{lnpr}	}&	\multicolumn{1}{.}{	-9.587	}&	\multicolumn{1}{.}{	0.775	}&	\multicolumn{1}{.}{	-28.572	}&	\multicolumn{1}{.}{	6.024	}&	\multicolumn{1}{c}{	$\sigma$	}&	\multicolumn{1}{.}{	29.572	}&	\multicolumn{1}{.}{	6.024	}\\ \cmidrule(r){1-8}
		\multicolumn{1}{l}{	obs.	}&	\multicolumn{2}{c}{	38	}&	\multicolumn{2}{c}{	4,718	}&	\multicolumn{3}{l}{		}\\												
\end{tabular*}																									
\begin{tabular*}{0.95\textwidth}{lrrrrrr..}																									
\multicolumn{9}{l}{--- Tests for 2SLS FE (IV) estimation ---\texttt{lnpfi pfi}}\\																						
	&	\multicolumn{5}{l}{	Underidentification}{Kleibergen-Paap rk LM statistic							}&	{		}&	\multicolumn{1}{.}{	27.145	}&	\multicolumn{1}{.}{	(0.000)	}\\						
	&	\multicolumn{5}{l}{	Weak identification}{Kleibergen-Paap rk Wald F statistic							}&	{		}&	\multicolumn{1}{.}{	15.968	}&	\multicolumn{1}{.}{		}\\						
	&	\multicolumn{5}{l}{	Overidentifying restriction}{Hansen J statistic							}&	{		}&	\multicolumn{1}{.}{	0.927	}&	\multicolumn{1}{.}{	(0.336)	}\\						
	&	\multicolumn{5}{l}{	Endogeneity}{statistic							}&	{		}&	\multicolumn{1}{.}{	14.055	}&	\multicolumn{1}{.}{	(0.000)	}\\						
\end{tabular*}																									
\begin{tabular*}{0.95\textwidth}{@{\extracolsep{\fill}}l}																									
\hline																									
\end{tabular*}																									
\begin{tablenotes}
\small
\item[Note 1] The numbers in parentheses for all tests are the p-values for rejecting the null hypotheses. 
\item[Note 2] Delta method estimates are based on FE (IV) according to the endogeneity test result.
\item[Note 3] All standard errors are robust to heteroskedasticity and autocorrelation.
\item[Note 4] Applied instruments: 1) exchange rates and 2) log of exchange rates.
\end{tablenotes}
\end{threeparttable}
\end{center}
\end{table}
\begin{table}[t!]
\caption{First-stage estimation for pork (prime)} \label{tab_4}\vspace{-10pt}
\begin{center}
\begin{threeparttable}
\begin{tabular*}{0.95\textwidth}{@{\extracolsep{\fill}}lccccrrr}
\hline\noalign{\smallskip}
\multicolumn{1}{l}{}&\multicolumn{2}{c}{FE (LS)}&\multicolumn{2}{c}{FE (IV)}&\multicolumn{3}{c}{Delta Method}\\\cmidrule(r){2-3}\cmidrule(r){4-5}\cmidrule(r){6-8}
&\multicolumn{1}{c}{coef.}&\multicolumn{1}{c}{s.e.}&\multicolumn{1}{c}{coef.}&\multicolumn{1}{c}{s.e.}&&\multicolumn{1}{c}{estim.}&\multicolumn{1}{c}{s.e.}\\\cmidrule(r){1-8}
\multicolumn{1}{l}{\texttt{lnpr}}&\multicolumn{1}{.}{-0.849}&\multicolumn{1}{.}{0.227}&\multicolumn{1}{.}{~~~~~~}&\multicolumn{1}{.}{~~~~~~}&\multicolumn{1}{c}{$\sigma$}&\multicolumn{1}{.}{1.849}&\multicolumn{1}{.}{0.227}\\ \cmidrule(r){1-8}
\multicolumn{1}{l}{obs.}&\multicolumn{2}{c}{74}&\multicolumn{2}{c}{}&\multicolumn{3}{l}{} 
\end{tabular*}
\begin{tabular*}{0.7\textwidth}{@{\extracolsep{\fill}}lllllllll}
\multicolumn{8}{l}{--- Test regression --- Dependent variable: \texttt{lnce}}\\
&\multicolumn{1}{l}{~~~~~~~~~~~~~~~}{}&\multicolumn{1}{c}{coef.}&\multicolumn{1}{c}{s.e.}&\multicolumn{1}{c}{}&\multicolumn{4}{l}{} \\ \cmidrule(r){1-8}
&\multicolumn{1}{l}{\texttt{lnjpylcc}}&\multicolumn{1}{.}{0.456}&\multicolumn{1}{.}{0.450}&\multicolumn{1}{.}{(0.314)}&\multicolumn{4}{l}{} \\
&\multicolumn{1}{l}{\texttt{cons}}&\multicolumn{1}{.}{8.367}&\multicolumn{1}{.}{0.284}&\multicolumn{1}{.}{(0.000)}&\multicolumn{4}{l}{} \\
\end{tabular*}
\begin{tabular*}{0.95\textwidth}{@{\extracolsep{\fill}}l}
\hline
\end{tabular*}

\begin{tablenotes}
\small
\item[Note 1] The numbers in parentheses for all tests are the p-values for rejecting the null hypotheses. 
\item[Note 2] All standard errors are robust to heteroskedasticity, except for the test regression.
\end{tablenotes}
\end{threeparttable}
\end{center}
\end{table}

In the second-stage estimation using the first-stage aggregates
obtained for pork (regular), our instruments do not appear to be relevant.
Thus, we investigated the possible feedback channel from $H$ of (\ref{secondreg2}) into ${R}$, which causes an endogeneity problem by letting the error term $\nu$ enter the explanatory variable $(R-Q)$.
If this channel is effective, $Q$ must also enter ${R}$, in which case $Q$ and $R$ become correlated.
Therefore, we regress $Q$ on $R$ in first differences, and find that
the standard error of the slope (i.e., 0.527), which yields a p-value of 0.068, suggests that the zero-slope hypothesis can not be rejected at the conventional level of significance.
We hence applied LS to obtain the estimates of the macroelasticity for the case of pork (regular), as presented in Table {\ref{tab_7}}.

\begin{table}[t!]
\caption{Second-stage estimation for pork (regular)} \label{tab_7}\vspace{-10pt}
\begin{center}
\begin{threeparttable}
\begin{tabular*}{0.95\textwidth}{@{\extracolsep{\fill}}lccccrrr}																									
\hline\noalign{\smallskip}																									
\multicolumn{1}{l}{}&\multicolumn{2}{c}{LS}&\multicolumn{2}{c}{IV}&\multicolumn{3}{c}{Delta Method}\\\cmidrule(r){2-3}\cmidrule(r){4-5}\cmidrule(r){6-8}																									
&\multicolumn{1}{c}{coef.}&\multicolumn{1}{c}{s.e.}&\multicolumn{1}{c}{coef.}&\multicolumn{1}{c}{s.e.}&&\multicolumn{1}{c}{estim.}&\multicolumn{1}{c}{s.e.}\\\cmidrule(r){1-8}																									
		\multicolumn{1}{l}{	\texttt{d.lnrq}	}&	\multicolumn{1}{.}{	0.504	}&	\multicolumn{1}{.}{	0.217	}&	\multicolumn{1}{.}{		}&	\multicolumn{1}{.}{		}&	\multicolumn{1}{c}{	$\rho$	}&	\multicolumn{1}{.}{	0.496	}&	\multicolumn{1}{.}{	0.217	}\\ \cmidrule(r){1-8}
		\multicolumn{1}{l}{	obs.	}&	\multicolumn{2}{c}{	24	}&	\multicolumn{2}{c}{		}&	\multicolumn{3}{l}{		}\\												
\end{tabular*}																									
\begin{tabular*}{0.7\textwidth}{@{\extracolsep{\fill}}lllllllll}																									
\multicolumn{8}{l}{--- Test regression --- Dependent variable: \texttt{d.lnpdi}}\\																									
&\multicolumn{1}{l}{~~~~~~~~~~~~~~~}{}&\multicolumn{1}{c}{coef.}&\multicolumn{1}{c}{s.e.}&\multicolumn{1}{c}{}&\multicolumn{4}{l}{} \\ \cmidrule(r){1-8}																									
&\multicolumn{1}{l}{\texttt{d.lnpfi}}&\multicolumn{1}{.}{1.012}&\multicolumn{1}{.}{0.527}&\multicolumn{1}{.}{(0.068)}&\multicolumn{4}{l}{} \\																									
&\multicolumn{1}{l}{\texttt{cons}}&\multicolumn{1}{.}{0.013}&\multicolumn{1}{.}{0.020}&\multicolumn{1}{.}{(0.524)}&\multicolumn{4}{l}{} \\																									
\end{tabular*}																									
\begin{tabular*}{0.95\textwidth}{@{\extracolsep{\fill}}l}																									
\hline																									
\end{tabular*}																									
\begin{tablenotes}
\small
\item[Note 1] The numbers in parentheses for all tests are the p-values for rejecting the null hypotheses. 
\item[Note 2] All standard errors are robust to heteroskedasticity, except for the test regression.
\end{tablenotes}
\end{threeparttable}
\end{center}
\end{table}

\section{Final Remarks}

To measure microelasticity, IV estimators were employed for beef,
chicken, and pork (regular), whereas we were not able to use the IV
estimator for pork (prime) because our IV was not sufficiently relevant.
Regarding macroelasticity, IV estimation was possible for beef and
chicken, although the LS estimator was employed for both cases owing
to the endogeneity test results, whereas this was not possible for
pork (regular) because our IV in this case was not sufficiently relevant.
In all cases, we further investigated the data to determine whether
there was reverse causality from the response variable to the
explanatory variable of the regression by examining the correlation
between the two components of the explanatory variable where either
could have had feedback effects from the response variable.
As we expected, such correlation was found in the first-stage
regression for beef, chicken, and pork (regular), where the endogeneity test result was positive and we hence employed the IV estimator, while such correlation was not found in the second-stage regression for beef and chicken, where the endogeneity test result was negative and we hence employed the LS estimator.
Since we did not find the abovementioned correlation in the remaining
regressions (i.e., the first stage for pork (prime) and second stage for
pork (regular)), we estimated the elasticities without being concerned about endogeneity in these cases.

Finally, we compare the results with those of previous studies.
Regarding microelasticity, recall that our point estimates were 4.35
(for beef), 4.01 (for chicken), 29.57 (for pork (regular)), and 1.85 (for
pork (prime)). 
These figures are comparable to 3.53 (for agriculture by \citet{saito}), 2.44 (for EU rice imports by \citet{jae2009}), 3.42 \citep{feenstra_restat18}, 4.01 (for beef by \citet{hide} via Feenstra's method), 17.32 (for chicken), and 33.00 (for pork).  
The GTAP microelasticities \citep{nber} are 7.70 (for bovine meat products) and 8.80 (for meat products etc.), which were derived by doubling the estimates of the macroelasticities, following the ``rule of two.''
We know that the large microelasticity for pork is due to the GPS
regime under which exporters have a strong incentive to set their price at the gate price regardless of the volume \citep{onji}.
Regarding macroelasticity, recall that our point estimates were 1.14
(for beef), 1.19 (for chicken), and 0.50 (for pork (regular)).
These figures are comparable to 1.68 (for meatpacking plants and
prepared meats by \citet{jpm92}), 1.07 (for chicken by \citet{kw}),
1.89 (for other livestock), 0.24 (for agriculture by \citet{saito}),
0.82 (for Japanese beef imports by \citet{kawashima}), 0.92 (for agriculture, forestry and fishing by \citet{baj2020}), 1.78 \citep{feenstra_restat18}, and 6.41 (for all meat by \citet{hide} via Feenstra's method). 
According to our elasticity estimates of import aggregation under a non-GPS regime (i.e., beef and chicken), it seems that a ``rule of four'' is more appropriate.

\def\thesection{Appendix}
\section{Summary of tariff scheme \label{appdx}}
\subsection*{Tariff duties for beef and chicken: 1996--2020}
\begin{figure}[t!]
\centering
\includegraphics[width=0.48\textwidth]{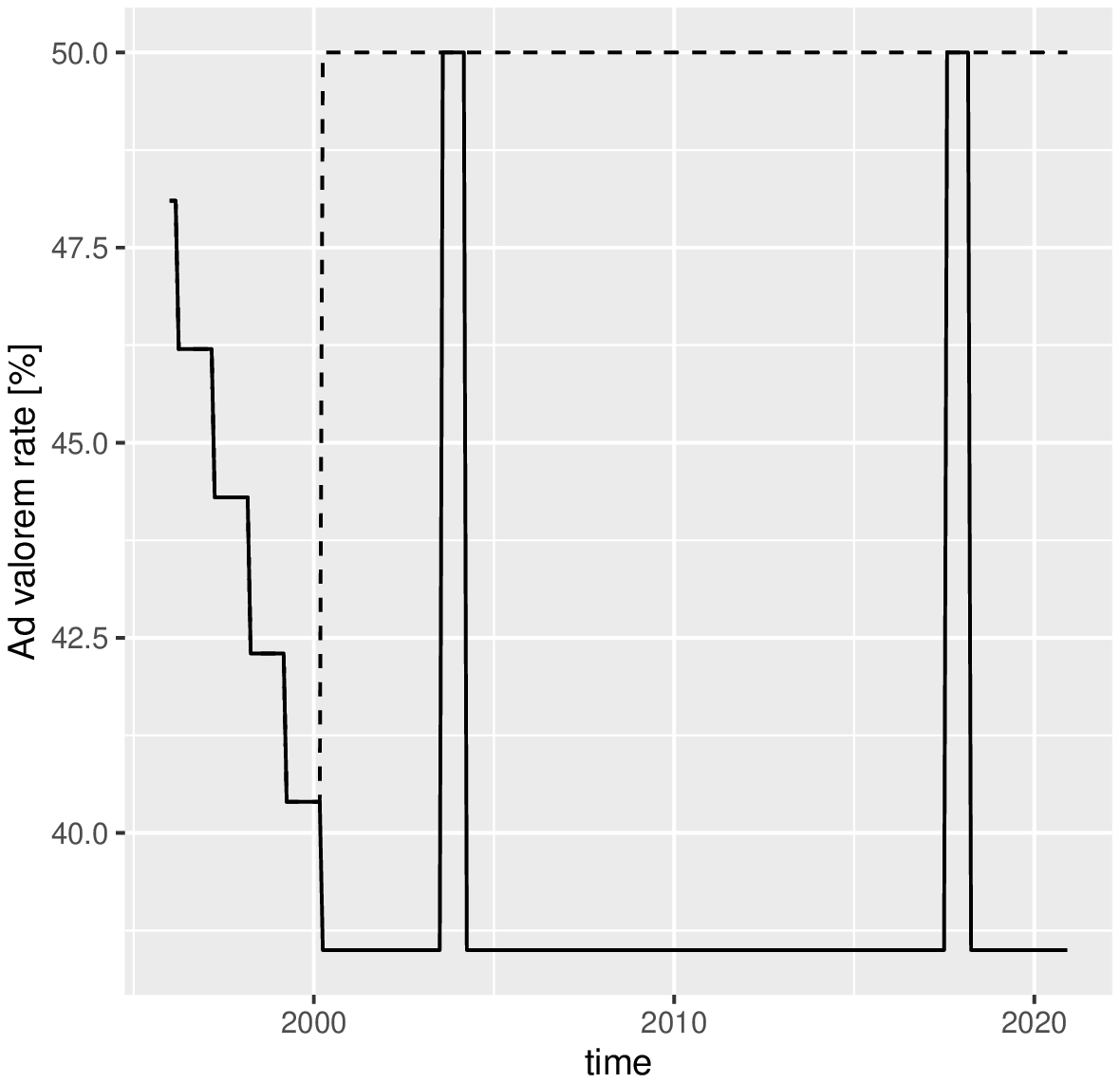}
\includegraphics[width=0.48\textwidth]{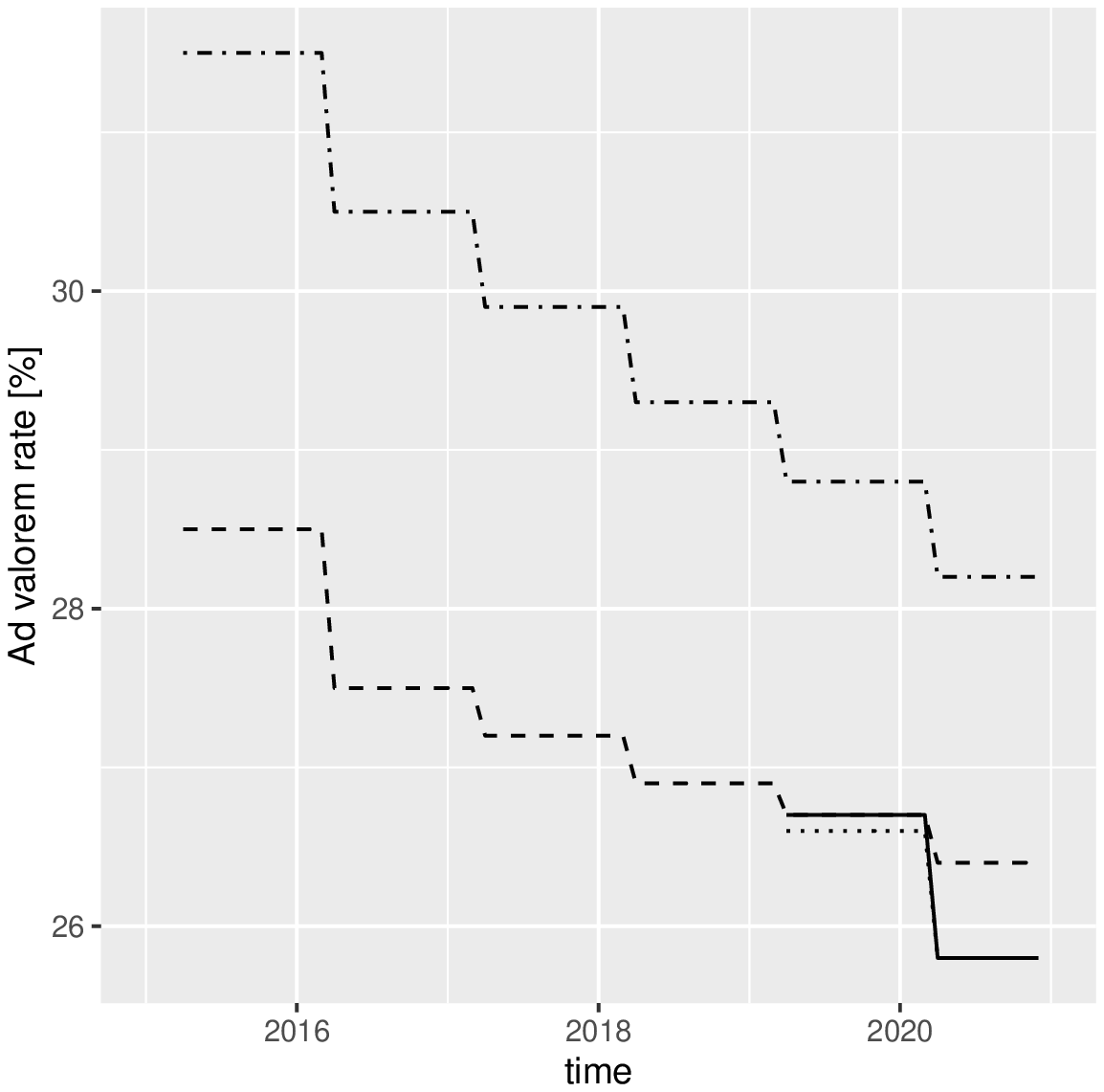}
\caption{
Left: Tariff rates applied to beef items from all countries except LDC
and those with an EPA (solid line).
General duty rate (dashed line).
Right: EPA tariff rates for Australia (dashed and dash-dotted lines), CPTPP (dotted line), and EU (solid line).
} \label{fig_beef}
\end{figure}

\begin{figure}[t!]
\centering
\includegraphics[width=0.48\textwidth]{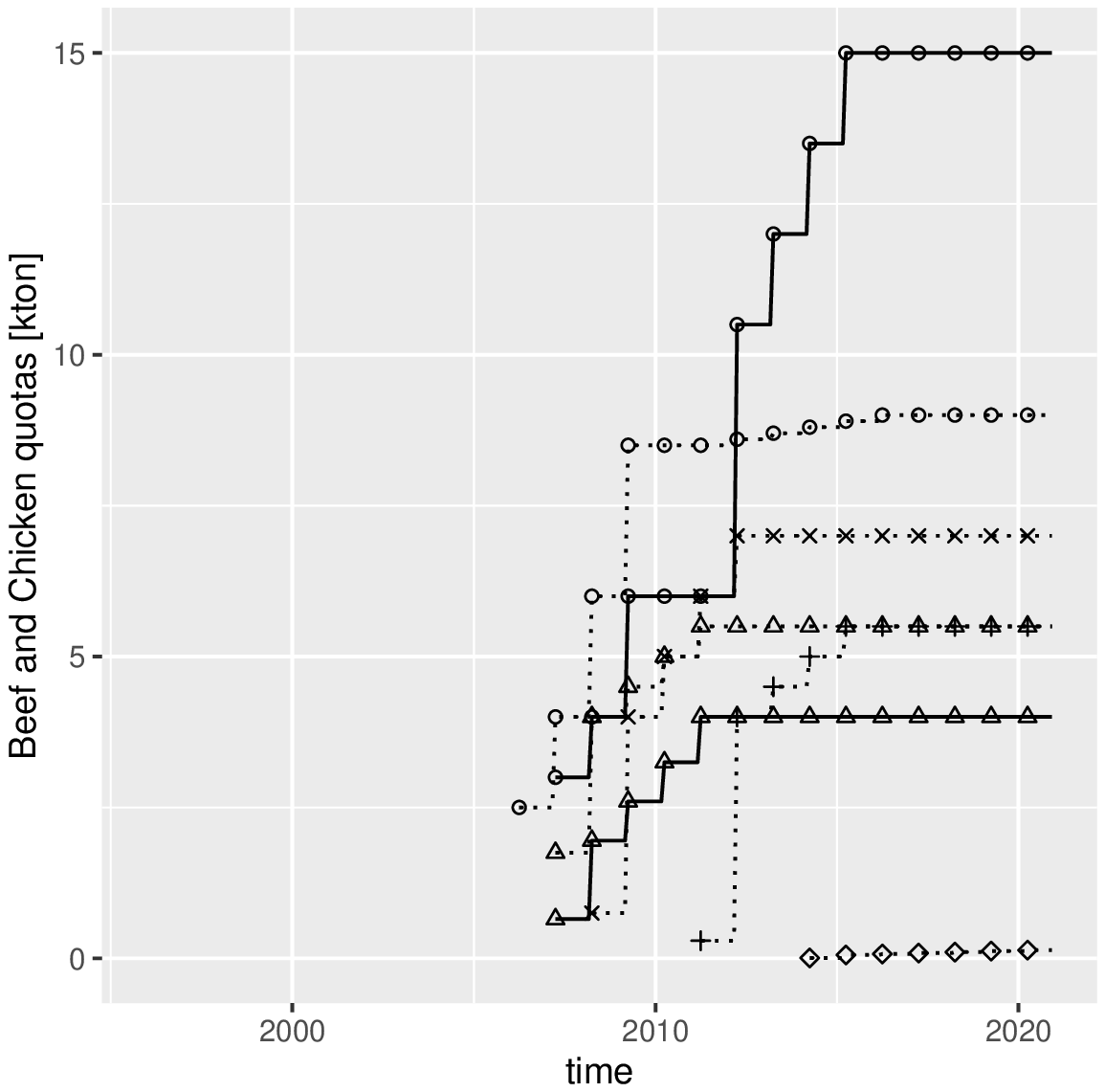}
\includegraphics[width=0.48\textwidth]{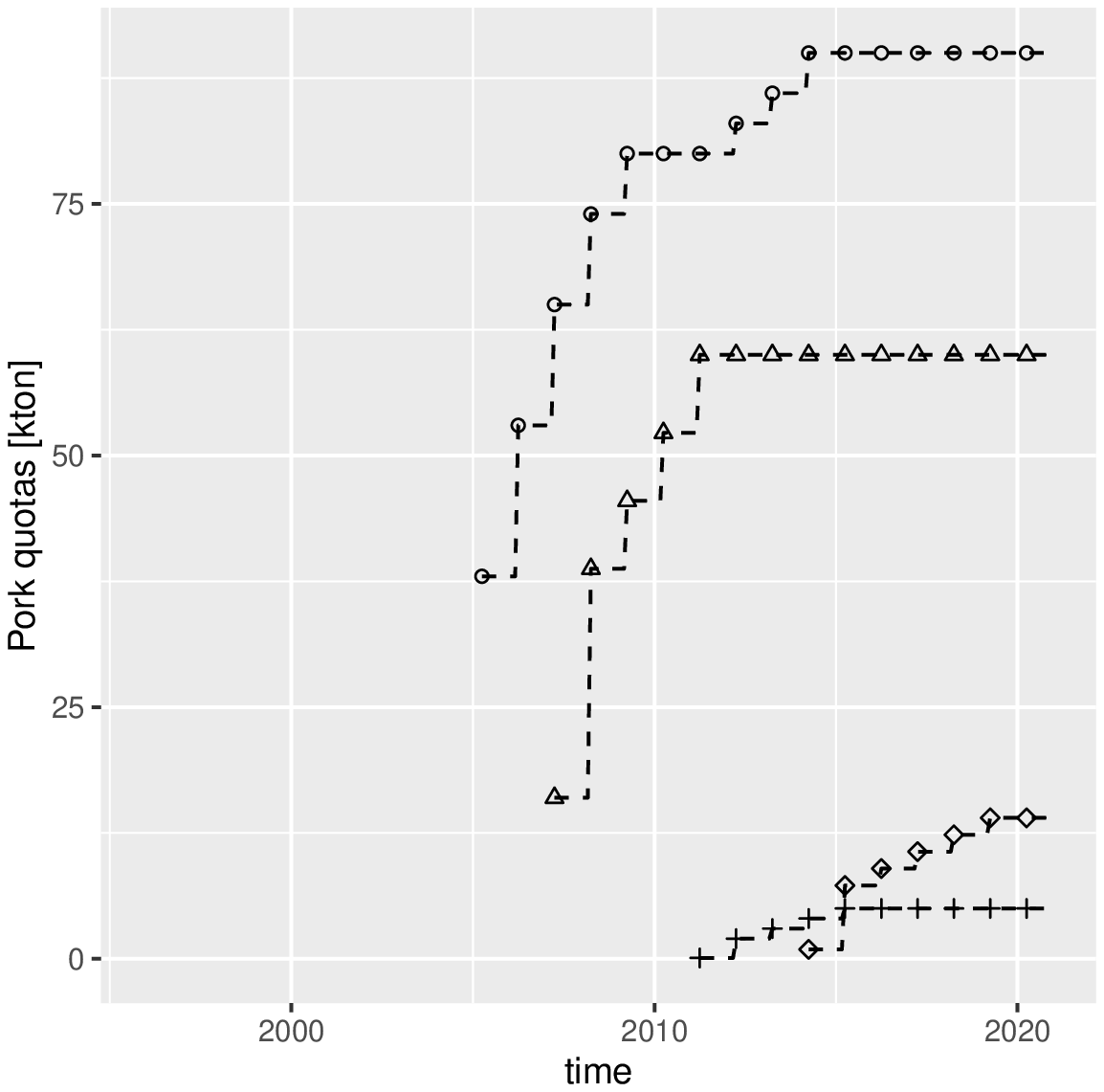}
\caption{Circles, triangles, pluses, crosses, and diamonds indicate Mexico (MEX), Chile (CHI), Peru (PER), The Philippines (PHI), and Australia (AUS), respectively.  
Solid, dashed, and dotted lines indicate beef, pork, and chicken, respectively.
} \label{fig_allquotas}
\end{figure}

Figure \ref{fig_beef}, left depicts the timeline of the ad valorem
tariff rate applied to all beef items, the HS codes of which are specified in Table \ref{tab_quotas}, except for those that were imported from LDC and countries with EPA, with a solid line. 
A dashed line in the same figure corresponds to a general duty rate,
which was applied to all beef items during the periods August 2003 -- March 2004 and August 2017 -- March 2018 (except for imports from Mexico, Chile, and Australia), due to safeguard activation.
All beef items imported from LDC have been subject to tariff exemption since JFY2007.\footnote{In Japan, the fiscal year runs from 1 April to 31 March of the following year. The Japanese fiscal year (JFY) from 1 April 2007 to 31 March 2008 would be abbreviated as JFY2007.} 
The EPA against Mexico since JFY2007 allows 34.6\% to be levied on
items 2 and 16 and 30.8\% to be levied on items 5--8, 10, and 13--15 for below-quota imports.
The EPA against Chile allowed 34.6\% to be levied on items 10 and 13--16 from JFY2007 to JFY2008 and 30.8\% on the same items since JFY2009 for below-quota imports. 
The timelines of annual quota limits for beef from Mexico and Chile are depicted in Figure \ref{fig_allquotas}, left, using solid lines with circles and triangles, respectively. 
The target items pertaining to quota limits against Mexico and Chile are specified in Table \ref{tab_quotas} with tag B.
Regarding the EPA against Australia, the tariff concession is
represented by the dashed line for items 1, 2, and 5--8 and the
dash-dotted line for items 9, 10, and 13--16 in Figure \ref{fig_beef} (right).
The EPA against EU and CPTPP countries allowed tariff concessions
depicted by the solid and dotted lines in Figure \ref{fig_beef}
(right), respectively.
A trade agreement with the US in JFY2020 allowed 25.8\% to be levied on all beef items from the US.

\begin{figure}[t!]
\centering
\includegraphics[width=0.48\textwidth]{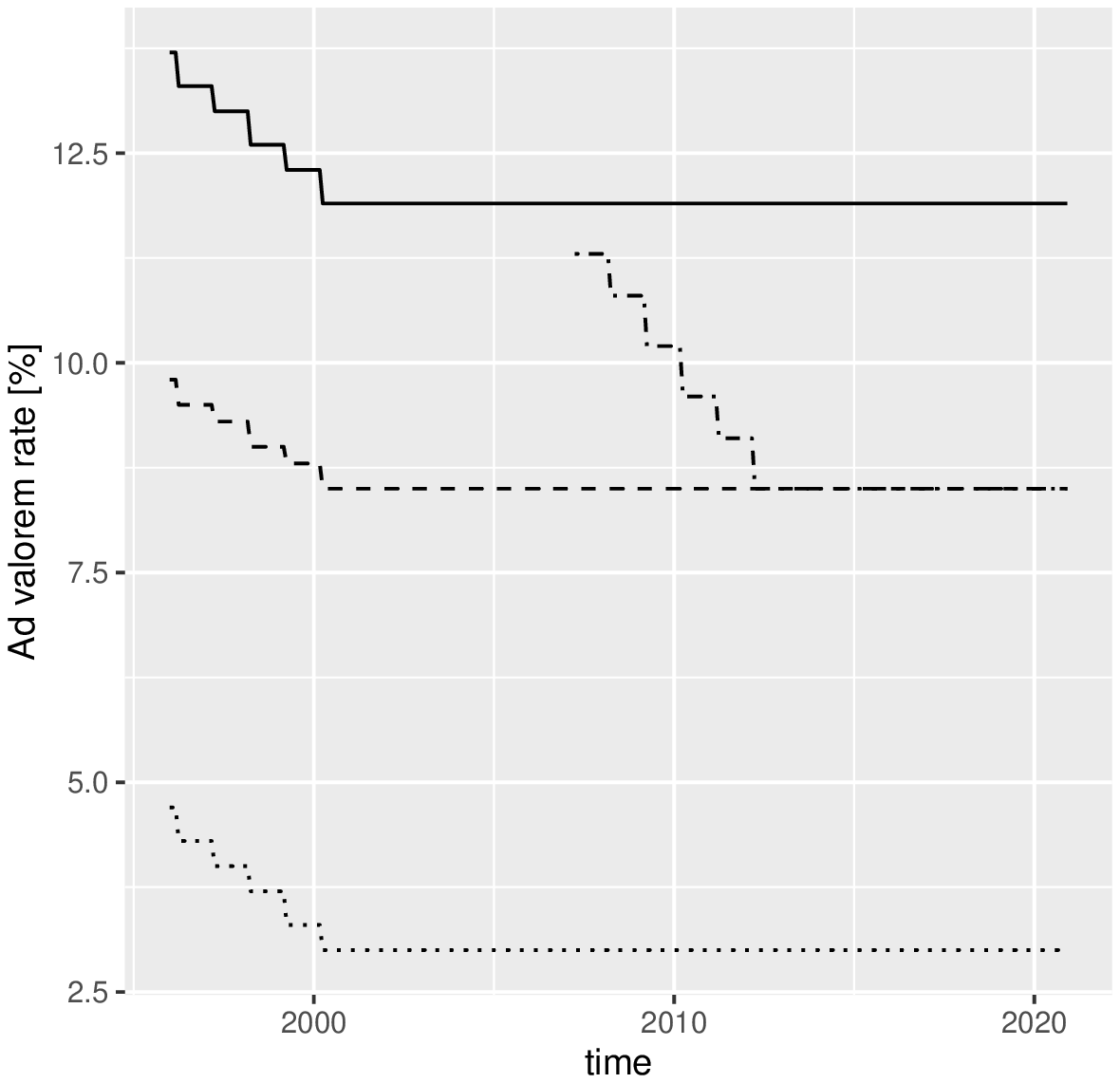}
\includegraphics[width=0.48\textwidth]{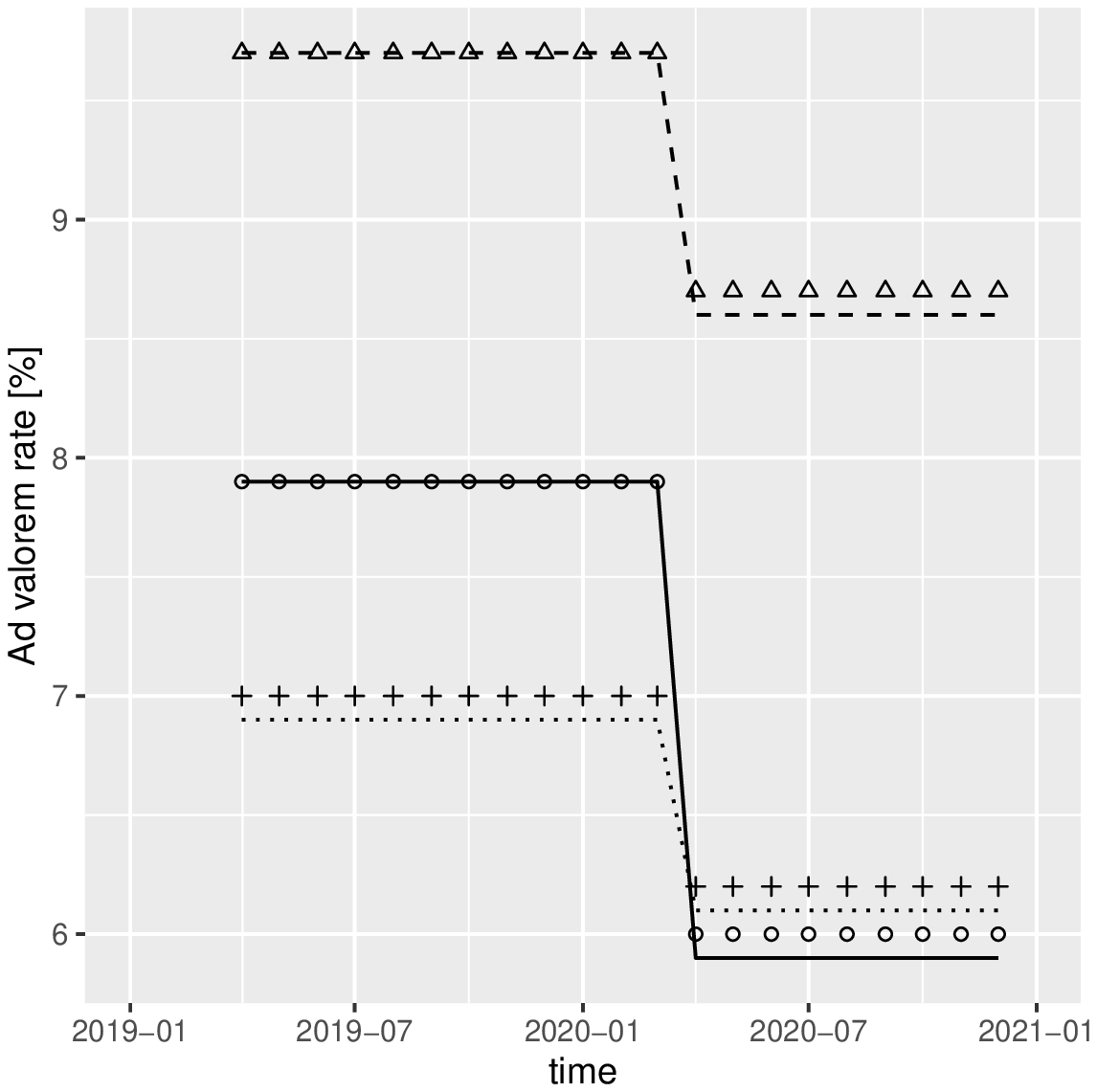}
\caption{
Left: Tariff rates applied to chicken items imported from MFN and ASEAN (since JFY2008) countries. 
Solid line corresponds to item groups a) and b).
Dashed line corresponds to item group c).
Dotted line corresponds to item group d).
The dot-dashed line corresponds to item groups a) and b) from Thailand since JFY2007. 
Right: Tariff rates applied to chicken items from CPTPP and EU countries.
Solid line and circles correspond to item group a) from CPTPP and EU countries, respectively.
Dashed line and triangles correspond to item group b) from CPTPP and EU countries, respectively.
Dotted line and pluses correspond to item group c) from CPTPP and EU countries, respectively.
} \label{fig_chicken}
\end{figure}

Regarding chicken items, the HS codes of which are specified in Table
\ref{tab_quotas}, the general tariff rate throughout the timeline was 14\% for items 68 and 69, 20\% for items 70 and 73, 12\% for items 70 and 73, and 10\% for item 72.
Tariff exemption has been granted on all chicken items from LDC and GSP countries. 
Tariff exemption has been also granted on all chicken items from Malaysia, Singapore, Indonesia, Brunei, Switzerland, Viet Nam, India, and Mongolia since JFY2006, JFY2008, JFY2008, JFY2008, JFY2010, JFY2010, JFY2011, and JFY2017, respectively.
Figure \ref{fig_chicken} left depicts the timeline of MFN tariff rates applied to item groups a) and b) with a solid line, c) with a dashed line, and d) with a dotted line, where we categorize the items into four groups i.e., a) 68 and 74, b) 69 and 71, c) 70 and 73, and d) 72.
The same figure includes the EPA tariff rate for chicken items 69, 71, and 74 from Thailand since JFY2007 with a dot-dashed line.
Since JFY2007, item 72 from Thailand has been subject to tariff exemption.
An ASEAN EPA that entered effect in JFY2008 allowed MFN tariff rates to be applied to all chicken items, except for item 72, which has been subject to tariff exemption.
Figure \ref{fig_chicken} right depicts the timeline of EPA tariff rates against CPTPP and EU countries since JFY2019.
Circles (solid line), triangles (dashed line), and pluses (dotted line) correspond to EU (CPTPP) tariff rates applied to chicken item groups a), b), and c), respectively.
Since JFY2019, item 72 from CPTPP and EU countries has been subject to tariff exemption.
A trade agreement with the US in JFY2020 allowed 8.6\%, 6.1\%, and 5.9\% on chicken items 69, 73, and 74, respectively, from the US. 

An EPA with Australia, since JFY2014, has allowed 10.7\% on items 68, 69, and 71, 7.6\% on item 70, tariff exemption on item 72, 6.8\% on item 73, and 8.5\% on item 74 for all below-quota imports.
The EPA with Peru, since JFY2011, has allowed 10.7\% on items 68, 69, 7.6\% on item 70, tariff exemption on item 72, 6.8\% on item 73, and 8.5\% (JFY2013 and after) and 10.7\% (before JFY2013) on item 74 for all below-quota imports.
The EPA with the Philippines, in place since JFY2008, allows 8.5\% on
items 68, 69, 71, and 74 and tariff exemption on item 72 for all below-quota imports.
The EPA with Chile, since JFY2007, allowed 8.5\% on item 74, and tariff exemption on item 72, for all below-quota imports.
Regarding the EPA with Mexico, the below-quota tariff rate for items
68, 69, and 71 was 10.7\% (from JFY2006 to JFY2011) and 7.1\% (from JFY2012), whereas that of item 70 was 7.6\% (from JFY2006 to JFY2011) and 5.1\% (from JFY2012).
For item 73, the below-quota tariff rate was 7.6\% (in JFY2006), 6.8\% (from JFY2007 to JFY2011), and 5.1\% (from JFY2012).
For item 74, the below-quota tariff rate was 10.7\% (in JFY2006), 8.5\% (from JFY2007 to JFY2011), and 7.1\% (from JFY2012).
The timeline of annual quota limits for chicken from Mexico, Chile, Peru, the Philippines, and Australia, is depicted in Figure \ref{fig_allquotas} left, using dotted lines with circles, triangles, pluses, crosses, and diamonds, respectively. 
The target items pertaining to quota limits with these five countries are specified in Table \ref{tab_quotas} with tag C.
\begin{figure}[t!]
\centering
\includegraphics[width=0.42\textwidth]{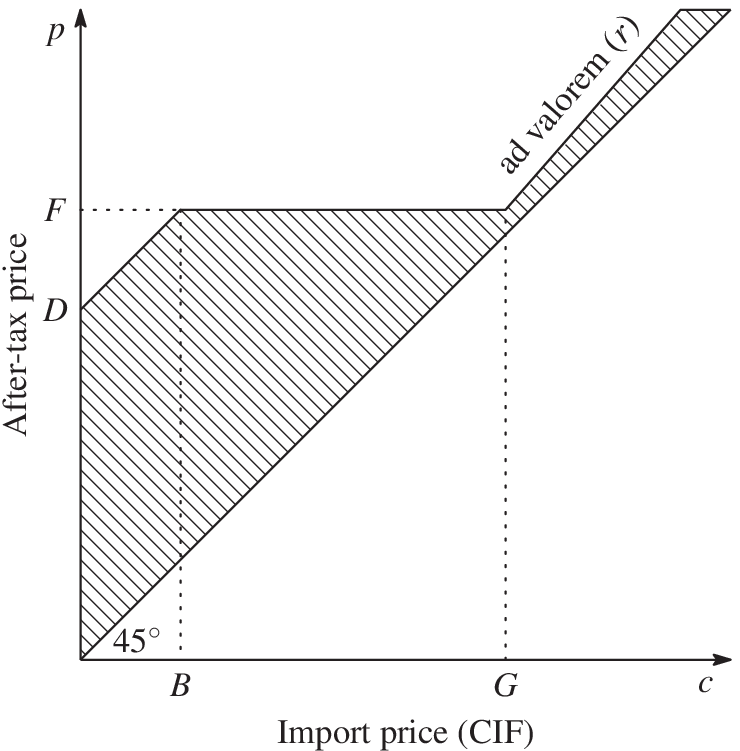}
~~~~~~
\includegraphics[width=0.48\textwidth]{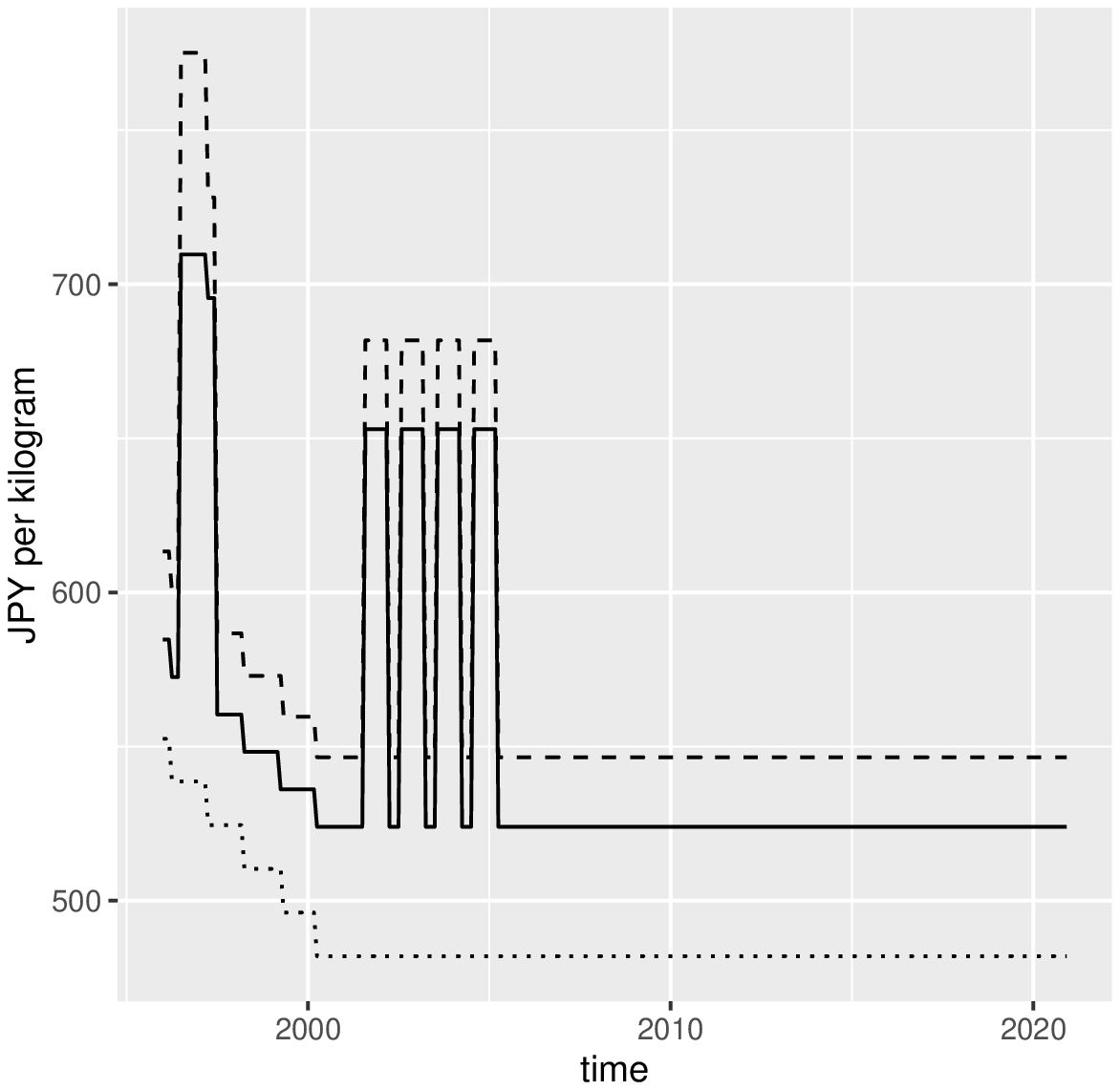}
\caption{Gate price system for Japan's pork imports.
Left: A unit tax ($D$) is applied if the import price ($c$) is below the threshold price ($B$).
Less duty will be levied if $c$ is between $B$ and the gate price ($G$), so that the after tax price ($p$) is leveled at the floor price ($F=D+B$).
If $c$ exceeds $G$, the greater of $rc$ and $F-c$ will be levied. 
The figure depicts the baseline case where $(1+r) G=F$.
Right: Solid, dashed, and dotted lines represent levels of $G$, $F$,
and $D$, respectively, on the timeline.
} \label{fig_gps}
\end{figure}

\subsection*{Tariff duties for pork: 1996--2020}
The tariff duty applied to pork imports to Japan has been subject to the GPS, which we explain with Figure \ref{fig_gps}.
GPS is a type of variable levy whereby the tariff duty depends on
import prices (of the CIF).
For each of the pork items shown in Table \ref{tab_quotas} IDs 28--48,
a per unit tax of $D$ applies if the import price $c$ is below the threshold price $B$.
If $c$ is above $B$ and below the gate price $G$, the levy is the difference between the floor price $F$ and $c$.
If $c$ is above $G$, the greater of the ad valorem tax of rate $r$ and the difference between $F$ and $c$ is applied.
We assume hereafter that the boundary values ($G, B, F, D$) of the GPS are those of non-carcass pork items (IDs 30--36, 39--38) in JPY per kilogram.
For carcass pork items (IDs 28--29, 37--38), these boundary values are reduced to 3/4, due to their meat content.
The boundary values $G, F$, and $D$, during the observation period (January 1996 -- December 2020), have changed, as depicted in Figure \ref{fig_gps}.
Note that the peaks observed during June 1996 -- March 1997 and August
2001 -- March 2005 were due to the activation of safeguards.
The baseline ad valorem tax rates ($r$) for all pork items have been
4.9\% (JFY1995), 4.8\% (JFY1996), 6.4\% (due to the activation of
safeguards in July 1996 --  March 1997 ), 4.7\% (JFY1997), 4.5\% (JFY1998), 4.4\% (JFY1999), and 4.3\% since JFY2000.

Japan's pork imports have been subject to quota limits on account of bilateral EPAs with Mexico, Chile, Peru, and Australia.
The target items pertaining to quota limits with these four countries are specified in Table \ref{tab_quotas} with tag P.
Reduced ad valorem tax rates and boundary values are applied to all pork items until the total volume of the target items exceeds the annual quota limit (which is shown in Figure \ref{fig_allquotas}).
Specifically, the sub-quota tax rates and boundary values are reduced
to $r=2.2$\%, $F=535.35$, $G=524$, and $D=482$ against Mexico, Chile,
Peru and Australia, while the baseline tax rates and boundary values
have been $r=4.3$\%, $F=546.35$, $G=524$, and $D=482$ since JFY2000.
In addition to the four EPA countries, Japan joined the CPTPP and
entered an FTA with the EU in JFY2019, thereby allowing reduced ad valorem rates ($r=1.9$\% for CPTPP and $r=2.0$\% for EU countries) and boundary values ($F=524\times 1.017$, $G=524$, $D=125$).
The same boundary values and a further reduced tax rate ($r=1.7$\%)
have been applied to items from CPTPP and EU countries and from the US
(via the Japan-US Trade Agreement) in JFY2020.

\begin{center}
\begin{ThreePartTable}
\begin{TableNotes}
\item[Note:] 
B, P, and C, indicate target items pertaining to beef, pork, and chicken, respectively.
\item[1] Items qualified for import quota with Mexico.
\item[2] Items qualified for import quota with Chile.
\item[3] Items qualified for import quota with Peru.
\item[4] Items qualified for import quota with the Philippines.
\item[5] Items qualified for import quota with Australia.
\end{TableNotes}
\begin{longtable}{clcccccccc}
\caption{Target items subject to import quotas.} \label{tab_quotas} \\
\hline\noalign{\smallskip}
\multicolumn{1}{c}{ID} & 
\multicolumn{1}{c}{HS code} &
\multicolumn{1}{c}{meat} &
\multicolumn{1}{c}{note1} &
\multicolumn{1}{c}{note2} &
\multicolumn{1}{c}{MEX$^{1}$} &
\multicolumn{1}{c}{CHI$^{2}$} &
\multicolumn{1}{c}{PER$^{3}$} &
\multicolumn{1}{c}{PHI$^{4}$} &
\multicolumn{1}{c}{AUS$^{5}$} \\
\hline\noalign{\smallskip}
\endfirsthead
\multicolumn{5}{c}%
{{\tablename\ \thetable{} -- continued from previous page}} \\
\hline\noalign{\smallskip}
\multicolumn{1}{c}{id} & 
\multicolumn{1}{c}{HS code} &
\multicolumn{1}{c}{meat} &
\multicolumn{1}{c}{note1} &
\multicolumn{1}{c}{note2} &
\multicolumn{1}{c}{MEX$^{1}$} &
\multicolumn{1}{c}{CHI$^{2}$} &
\multicolumn{1}{c}{PER$^{3}$} &
\multicolumn{1}{c}{PHI$^{4}$} &
\multicolumn{1}{c}{AUS$^{5}$} \\
 \hline\noalign{\smallskip}
\endhead
\hline
\endfoot
\hline 
\insertTableNotes
\endlastfoot
1	&	0201.10-000	&	Beef	&	fresh	&	carcass	&		&		&		&		&		\\
2	&	0201.20-000	&	Beef	&	fresh	&	bonein	&	{B}	&		&		&		&		\\
3	&	0201.20-010	&	Beef	&	fresh	&	bonein	&		&		&		&		&		\\
4	&	0201.20-090	&	Beef	&	fresh	&	bonein	&		&		&		&		&		\\
5	&	0201.30-010	&	Beef	&	fresh	&	boneless	&	{B}	&		&		&		&		\\
6	&	0201.30-020	&	Beef	&	fresh	&	boneless	&	{B}	&		&		&		&		\\
7	&	0201.30-030	&	Beef	&	fresh	&	boneless	&	{B}	&		&		&		&		\\
8	&	0201.30-090	&	Beef	&	fresh	&	boneless	&	{B}	&		&		&		&		\\
9	&	0202.10-000	&	Beef	&	freezed	&	carcass	&		&		&		&		&		\\
10	&	0202.20-000	&	Beef	&	freezed	&	bonein	&	{B}	&	{B}	&		&		&		\\
11	&	0202.20-010	&	Beef	&	freezed	&	bonein	&		&		&		&		&		\\
12	&	0202.20-090	&	Beef	&	freezed	&	bonein	&		&		&		&		&		\\
13	&	0202.30-010	&	Beef	&	freezed	&	boneless	&	{B}	&	{B}	&		&		&		\\
14	&	0202.30-020	&	Beef	&	freezed	&	boneless	&	{B}	&	{B}	&		&		&		\\
15	&	0202.30-030	&	Beef	&	freezed	&	boneless	&	{B}	&	{B}	&		&		&		\\
16	&	0202.30-090	&	Beef	&	freezed	&	boneless	&	{B}	&	{B}	&		&		&		\\
17	&	0206.10-010	&		&		&		&	{B}	&		&		&		&		\\
18	&	0206.21-000	&		&		&		&	{B}	&		&		&		&		\\
19	&	0206.22-000	&		&		&		&	{B}	&		&		&		&		\\
20	&	0206.29-010	&		&		&		&	{B}	&		&		&		&		\\
21	&	0206.29-020	&		&		&		&	{B}	&		&		&		&		\\
22	&	0206.29-090	&		&		&		&	{B}	&		&		&		&		\\
23	&	1602.50-510	&		&		&		&	{B}	&		&		&		&		\\
24	&	1602.50-520	&		&		&		&	{B}	&		&		&		&		\\
25	&	1602.50-590	&		&		&		&	{B}	&		&		&		&		\\
26	&	1602.50-600	&		&		&		&	{B}	&		&		&		&		\\
27	&	1602.50-700	&		&		&		&	{B}	&		&		&		&		\\
28	&	0203.11-030	&	Pork	&	fresh	&	carcass	&		&		&		&		&	{P}	\\
29	&	0203.11-040	&	Pork	&	fresh	&	carcass	&		&		&		&		&	{P}	\\
30	&	0203.12-021	&	Pork	&	fresh	&	bonein	&	{P}	&		&		&		&	{P}	\\
31	&	0203.12-022	&	Pork	&	fresh	&	bonein	&	{P}	&		&		&		&	{P}	\\
32	&	0203.12-025	&	Pork	&	fresh	&	bonein	&	{P}	&		&		&		&	{P}	\\
33	&	0203.19-021	&	Pork	&	fresh	&	boneless	&	{P}	&	{P}	&		&		&	{P}	\\
34	&	0203.19-022	&	Pork	&	fresh	&	boneless	&	{P}	&	{P}	&		&		&	{P}	\\
35	&	0203.19-024	&	Pork	&	fresh	&	boneless	&	{P}	&	{P}	&		&		&	{P}	\\
36	&	0203.19-025	&	Pork	&	fresh	&	boneless	&	{P}	&	{P}	&		&		&	{P}	\\
37	&	0203.21-030	&	Pork	&	freezed	&	carcass	&		&		&		&		&	{P}	\\
38	&	0203.21-040	&	Pork	&	freezed	&	carcass	&		&		&		&		&	{P}	\\
39	&	0203.22-021	&	Pork	&	freezed	&	bonein	&	{P}	&	{P}	&	{P}	&		&	{P}	\\
40	&	0203.22-022	&	Pork	&	freezed	&	bonein	&	{P}	&	{P}	&	{P}	&		&	{P}	\\
41	&	0203.22-023	&	Pork	&	freezed	&	bonein	&	{P}	&	{P}	&	{P}	&		&	{P}	\\
42	&	0203.22-024	&	Pork	&	freezed	&	bonein	&	{P}	&	{P}	&	{P}	&		&	{P}	\\
43	&	0203.22-025	&	Pork	&	freezed	&	bonein	&	{P}	&	{P}	&	{P}	&		&	{P}	\\
44	&	0203.29-021	&	Pork	&	freezed	&	boneless	&	{P}	&	{P}	&	{P}	&		&	{P}	\\
45	&	0203.29-022	&	Pork	&	freezed	&	boneless	&	{P}	&	{P}	&	{P}	&		&	{P}	\\
46	&	0203.29-023	&	Pork	&	freezed	&	boneless	&	{P}	&	{P}	&	{P}	&		&	{P}	\\
47	&	0203.29-024	&	Pork	&	freezed	&	boneless	&	{P}	&	{P}	&	{P}	&		&	{P}	\\
48	&	0203.29-025	&	Pork	&	freezed	&	boneless	&	{P}	&	{P}	&	{P}	&		&	{P}	\\
49	&	0206.49-091	&		&		&		&		&		&		&		&	{P}	\\
50	&	0206.49-092	&		&		&		&	{P}	&	{P}	&		&		&		\\
51	&	0206.49-094	&		&		&		&	{P}	&	{P}	&		&		&		\\
52	&	0206.49-099	&		&		&		&	{P}	&	{P}	&		&		&		\\
53	&	0210.11-010	&		&		&		&	{P}	&	 	&		&		&		\\
54	&	0210.11-020	&		&		&		&	{P}	&		&		&		&		\\
55	&	0210.12-010	&		&		&		&	{P}	&		&		&		&		\\
56	&	0210.12-020	&		&		&		&	{P}	&		&		&		&		\\
57	&	0210.19-010	&		&		&		&	{P}	&		&		&		&		\\
58	&	0210.19-020	&		&		&		&	{P}	&		&		&		&		\\
59	&	1602.41-011	&		&		&		&	{P}	&	{P}	&		&		&	{P}	\\
60	&	1602.41-019	&		&		&		&	{P}	&	{P}	&		&		&	{P}	\\
61	&	1602.41-090	&		&		&		&		&	{P}	&		&		&	{P}	\\
62	&	1602.42-011	&		&		&		&	{P}	&	{P}	&		&		&	{P}	\\
63	&	1602.42-019	&		&		&		&	{P}	&	{P}	&		&		&	{P}	\\
64	&	1602.42-090	&		&		&		&		&	{P}	&		&		&	{P}	\\
65	&	1602.49-210	&		&		&		&	{P}	&	{P}	&		&		&	{P}	\\
66	&	1602.49-220	&		&		&		&	{P}	&	{P}	&		&		&	{P}	\\
67	&	1602.49-290	&		&		&		&		&	{P}	&		&		&	{P}	\\
68	&	0207.11-000	&	Chicken	&	gallus d.	&		&	{C}	&		&	{C}	&	{C}	&	{C}	\\
69	&	0207.12-000	&	Chicken	&	gallus d.	&		&	{C}	&		&	{C}	&	{C}	&	{C}	\\
70	&	0207.13-100	&	Chicken	&	gallus d.	&		&	{C}	&		&	{C}	&		&	{C}	\\
71	&	0207.13-200	&	Chicken	&	gallus d.	&		&	{C}	&		&		&	{C}	&	{C}	\\
72	&	0207.14-100	&	Chicken	&	gallus d.	&	liver	&		&		&		&		&		\\
73	&	0207.14-210	&	Chicken	&	gallus d.	&		&	{C}	&		&	{C}	&		&	{C}	\\
74	&	0207.14-220	&	Chicken	&	gallus d.	&		&	{C}	&	{C}	&	{C}	&	{C}	&	{C}	\\
75	&	1602.31-210	&		&		&		&	{C}	&		&	{C}	&		&	{C}	\\
76	&	1602.32-210	&		&		&		&	{C}	&		&	{C}	&		&	{C}	\\
77	&	1602.32-290	&		&		&		&	{C}	&		&	{C}	&		&	{C}	\\
78	&	1602.39-210	&		&		&		&	{C}	&		&	{C}	&		&	{C}	\\

\end{longtable}
\end{ThreePartTable}
\end{center}

\section*{Acknowledgements}
JSPS Kakenhi Grant numbers: 19H04380, 20K22139 \\
The authors declare that they have no conflicts of interest.

{\raggedright
\bibliography{bibfile}
}

\end{document}